\documentclass[aps,prd,amsmath,amssymb,superscriptaddress,twocolumn,nofootinbib,showpacs,preprintnumbers,notitlepage]{revtex4-1}
\usepackage[usenames,dvipsnames,svgnames,table]{xcolor}
\usepackage{graphicx}
\usepackage{dcolumn}
\usepackage{bm}
\usepackage{hyperref}
\hypersetup{ 
    pdfnewwindow=true,      
    colorlinks=true,       
    allcolors=[RGB]{31 119 180}
} 
\usepackage[capitalise]{cleveref}
\usepackage{dsfont}
\usepackage{placeins}
\usepackage{todonotes}
\usepackage[utf8]{inputenc}
\usepackage{gensymb}
\usepackage{multirow}
\usepackage{amsmath}
\usepackage{amsfonts}
\usepackage{amssymb}
\usepackage{enumitem,amssymb}
\usepackage{array}   
\newcolumntype{L}{>{$}l<{$}} 
\newcolumntype{R}{>{$}r<{$}}
\newcolumntype{C}{>{$}c<{$}}
\usepackage{xspace}
\usepackage{braket}
\usepackage{units}
\usepackage{slashed}
\usepackage[normalem]{ulem}
\usepackage[format=plain,justification=RaggedRight,singlelinecheck=false]{caption}
\usepackage[format=plain,justification=centering,singlelinecheck=false]{subcaption}
\usepackage{tabularx}
\usepackage{nameref}

\usepackage{xpatch}
\makeatletter
\xpatchcmd{\@ssect@ltx}{\@xsect}{\protected@edef\@currentlabelname{#8}\@xsect}{}{}
\xpatchcmd{\@sect@ltx}{\@xsect}{\protected@edef\@currentlabelname{#8}\@xsect}{}{}
\makeatother
\usepackage{hyperref}

\graphicspath{{figures/}}

\let\Re\relax
\let\Im\relax
\DeclareMathOperator{\Re}{Re}
\DeclareMathOperator{\Im}{Im}
\DeclareMathOperator{\sign}{sign}

\newcommand{\eg}{{\it e.g.}\xspace}

\newcommand{\ie}{{\it i.e.}\xspace}

\newcommand{\mevnospace}{\ensuremath{{\mathrm{\,Me\kern -0.1em V}}}}
\newcommand{\gevnospace}{\ensuremath{{\mathrm{\,Ge\kern -0.1em V}}}}
\newcommand{\tevnospace}{\ensuremath{{\mathrm{\,Te\kern -0.1em V}}}}

\newcommand\bsub{\begin{subequations}}
\newcommand\esub{\end{subequations}}
\newcommand{\diff}{\ensuremath{\mathrm{d}}}

\newlist{todolist}{itemize}{2}
\setlist[todolist]{label=$\square$}
\usepackage{pifont}

\usepackage{mfirstuc} 
\usepackage{todonotes} 
\newcommand{\addReviewer}[2]{
  \expandafter\newcommand\csname #1\endcsname[1]{{\textbf{ \color{#2} \capitalisewords{#1}:\,##1}}}
  \expandafter\newcommand\csname #1cor\endcsname[2]{{\color{#2} \capitalisewords{#1}:\,\st{##1}{\textbf{##2}}}}
  \expandafter\newcommand\csname #1color\endcsname{#2}
  \expandafter\newcommand\csname #1todo\endcsname[1]{{\todo[inline,color=white!70!#2, caption={}]{\textbf{\capitalisewords{#1}}: ##1}}}
}

\usepackage{tikz, marginnote} 
\newcommand{\checkedby}[1]{
\ifdefined\CROSSCHECKS
  \marginnote{
    \begin{tikzpicture}
      \foreach \x [count=\xi] in {#1} {
         \node[shape=circle,inner sep=0mm,
         minimum size=2mm,
         fill=\csname \x color\endcsname] at (\xi*3mm,0) {};
       }
    \end{tikzpicture}
  }
\else
\fi
}

\usepackage{soul,color}

\definecolor{chromeyellow}{rgb}{1.0, 0.65, 0.0}
\definecolor{DodgeBlue}{rgb}{0.118, 0.565,1.000}
\definecolor{asparagus}{rgb}{0.53, 0.66, 0.42}
\definecolor{cadmiumgreen}{rgb}{0.0, 0.42, 0.24}
\definecolor{blue(ryb)}{rgb}{0.01, 0.28, 1.0}
\definecolor{periwinkle}{RGB}{181, 146, 203}
\definecolor{turquoiseblue}{rgb}{0.02, 0.55, 0.55}
\definecolor{green1}{RGB}{50,205,50}
\definecolor{amethyst}{rgb}{0.6, 0.4, 0.8}
\definecolor{indianred}{RGB}{205,92,92}
\definecolor{applegreen}{rgb}{0.55,0.71,0.0}
\definecolor{teal}{RGB}{66, 245, 197}

\addReviewer{bernhard}{orange}
\addReviewer{adam}{red}
\addReviewer{ale}{chromeyellow}
\addReviewer{vincent}{brown}
\addReviewer{misha}{cadmiumgreen}
\addReviewer{cesar}{magenta}
\addReviewer{tomasz}{cyan}
\addReviewer{miguel}{DodgeBlue}
\addReviewer{andrew}{purple}
\addReviewer{arkaitz}{asparagus}
\addReviewer{yannick}{blue(ryb)}
\addReviewer{mathias}{turquoiseblue}
\addReviewer{lukasz}{blue}
\addReviewer{daniel}{teal}
\addReviewer{robert}{turquoiseblue}
\addReviewer{andreac}{orange}
\addReviewer{igor}{green1}
\addReviewer{sergi}{Orchid}
\addReviewer{astrid}{ForestGreen}
\addReviewer{gloria}{indianred}
\addReviewer{wyatt}{amethyst}
\addReviewer{derek}{purple}
\addReviewer{zach}{applegreen}
\addReviewer{vanamali}{periwinkle}

\usepackage{orcidlink}

\newcommand{\catania}{INFN Sezione di Catania, I-95123 Catania, Italy}
\newcommand{\ceem}{Center for  Exploration  of  Energy  and  Matter, Indiana  University, Bloomington,  IN  47403,  USA}
\newcommand{\icn}{Instituto de Ciencias Nucleares, Universidad Nacional Aut\'onoma de M\'exico, Ciudad de M\'exico 04510, Mexico}
\newcommand{\ific}{Instituto de F\'isica Corpuscular (IFIC), Centro Mixto CSIC-Universidad de Valencia, E-46071 Valencia, Spain}
\newcommand{\ifj}{Institute of Nuclear Physics, Polish Academy of Sciences, PL-31-342 Krak\'ow, Poland}
\newcommand{\indiana}{Department of Physics, Indiana  University, Bloomington,  IN  47405,  USA}
\newcommand{\jlab}{Theory Center, Thomas  Jefferson  National  Accelerator  Facility, Newport  News,  VA  23606,  USA}
\newcommand{\jlabNonTheo}{Thomas  Jefferson  National  Accelerator  Facility, Newport  News,  VA  23606,  USA}

\newcommand{\lmu}{Ludwig-Maximilians-Universit{\"a}t, D-80539 Munich, Germany}
\newcommand{\messina}{Dipartimento di Scienze Matematiche e Informatiche, Scienze Fisiche e Scienze della Terra, Universit\`a degli Studi di Messina, I-98122 Messina, Italy}
\newcommand{\origins}{ORIGINS Excellence Cluster, D-80939 Munich, Germany}
\newcommand{\scnuIQM}{Guangdong Provincial Key Laboratory of Nuclear Science, Institute of Quantum Matter, South China Normal University, Guangzhou 510006, China}
\newcommand{\scnuJLQM}{Guangdong-Hong Kong Joint Laboratory of Quantum Matter, Southern Nuclear Science Computing Center, South China Normal University, Guangzhou 510006, China}
\newcommand{\ub}{Departament de F\'isica Qu\`antica i Astrof\'isica and Institut de Ci\`encies del Cosmos, Universitat de Barcelona,  E-08028 Barcelona, Spain}
\newcommand{\ucm}{Departamento de F\'isica Te\'orica, Universidad Complutense de Madrid and IPARCOS, E-28040 Madrid, Spain}
\newcommand{\uned}{Departamento de F\'isica Interdisciplinar, Universidad Nacional de Educaci\'on a Distancia (UNED), E-28040 Madrid, Spain}

\newcommand{\glasgow}{School of Physics and Astronomy, University of Glasgow, Glasgow, G12 8QQ, UK}
\newcommand{\cmu}{Department of Physics, Carnegie Mellon University, Pittsburgh, Pennsylvania 15213, USA}  

\begin{document}

\preprint{JLAB-THY-23-3873}
\title{ Ambiguities in Partial Wave Analysis of Two Spinless Meson Photoproduction}

\author{W.~A.~\surname{Smith}\orcidlink{0009-0001-3244-6889}}
\email{smithwya@iu.edu}
\affiliation{\indiana}
\affiliation{\ceem}

\author{D.~I.~\surname{Glazier}\orcidlink{0000-0002-8929-6332}}
\affiliation{\glasgow}

\author{V.~\surname{Mathieu}\orcidlink{0000-0003-4955-3311}}
\affiliation{\ub}
\affiliation{\ucm}

\author{M.~\surname{Albaladejo}\orcidlink{0000-0001-7340-9235}}
\affiliation{\ific}

\author{M.~\surname{Albrecht}\orcidlink{0000-0001-6180-4297}}
\affiliation{\jlabNonTheo}

\author{Z.~\surname{Baldwin}\orcidlink{0000-0002-8534-0922}}
\affiliation{\cmu}

\author{C.~\surname{Fern\'andez-Ram\'irez}\orcidlink{0000-0001-8979-5660}}
\affiliation{\uned}
\affiliation{\icn}

\author{N.~\surname{Hammoud}\orcidlink{0000-0002-8395-0647}}
\affiliation{\ifj}

\author{M.~\surname{Mikhasenko}\orcidlink{0000-0002-6969-2063}}
\affiliation{\origins}
\affiliation{\lmu}

\author{G.~\surname{Monta\~na}\orcidlink{0000-0001-8093-6682}}
\affiliation{\jlab}

\author{R.~J.~\surname{Perry}\orcidlink{0000-0002-2954-5050}}
\affiliation{\ub}

\author{A.~\surname{Pilloni}\orcidlink{0000-0003-4257-0928}}
\affiliation{\messina}
\affiliation{\catania}

\author{V.~\surname{Shastry}\orcidlink{0000-0003-1296-8468}}
\affiliation{\indiana}
\affiliation{\ceem}

\author{A.~P.~\surname{Szczepaniak}\orcidlink{0000-0002-4156-5492}}
\affiliation{\indiana}
\affiliation{\ceem}
\affiliation{\jlab}

\author{D.~\surname{Winney}\orcidlink{0000-0002-8076-243X}}
\affiliation{\scnuIQM}
\affiliation{\scnuJLQM}

\collaboration{Joint Physics Analysis Center}
\begin{abstract}
We 
describe the formalism 
to analyze the mathematical ambiguities arising in partial-wave analysis of two spinless mesons produced with a linearly polarized photon beam. We show that partial waves are uniquely defined when all accessible observables are considered, for a wave set which includes $S$ and $D$ waves. The inclusion of higher partial waves does not affect our results, and we conclude that there are no mathematical ambiguities in partial-wave analysis of two mesons produced with a linearly polarized photon beam.
We present Monte Carlo simulations to illustrate our results. 
\end{abstract}
\maketitle

\section{Introduction}
\label{sec:intro}

In hadron spectroscopy, the extraction and interpretation of data from scattering experiments typically employ partial-wave analyses to isolate resonant contributions.
However, these partial-wave expansions need not be unique, and, depending on the reaction, one may find multiple wave sets which produce mathematically equivalent predictions for the observables. This causes significant problems in the analysis and interpretation of data.
These mathematical ambiguities have been extensively studied for various 
processes~\cite{Barrelet:1971pw,Chung:1997qd,Sadovsky:1999gt,Austregesilo:2014oxa} and there is no generic prescription to remedy them. Hence, the issue must be addressed on a case-by-case basis (see Refs.~\cite{Ketzer:2019wmd, Rodas:2021tyb, Gao:2023jtq,
Kroenert:2023ovd} for some recent examples).
To remedy ambiguities, typically one must generate all possible ambiguous wave sets and select one of them by enforcing additional constraints like global continuity~\cite{Kok:1976} or unitarity~\cite{BESIII:2015rug}. Most previous analyses of mathematical ambiguities for partial-wave analysis examine nucleon or pion-beam production processes. In this work, we introduce the formalism for the examination of mathematical ambiguities in two pseudoscalar meson photoproduction processes with a linearly polarized photon beam, such as those present in the GlueX experiment at Jefferson Lab~\cite{Mathieu:2019fts}.

The physics program for the GlueX experiment focuses on the search for light exotic mesons. Some of the final states under consideration involve the two pseudoscalar mesons $\eta^{(\prime)} \pi$, for which odd waves have exotic quantum numbers incompatible with a $q\bar q$ assignment~\cite{Meyer:2010ku}. The dominant non-exotic signal in these final states is the  $a_2(1320)$ resonance which populates the $D$ waves~\cite{CLAS:2020rdz}. 
It is essential to first accurately identify all relevant $D$-wave components before extracting the weaker exotic signal in the $P$-waves~\cite{Ketzer:2019wmd,JPAC:2018zyd,Kopf:2020yoa,Woss:2020ayi}. In this paper, we address the issue of ambiguous solutions in partial wave analyses which are relevant to the extraction of the $D$-wave components, but our work is applicable to the general case of photoproproduction of any two spinless mesons.
Our methods are based on the concept of Barrelet zeros, which we review in \cref{sec:Barrelet} for completeness. In \cref{sec:formalism} we introduce our notation and formalism for the photoproduction of two spinless mesons with a linearly polarized photon beam. We then demonstrate, using a wave set with two or three $D$-wave components accompanied by an $S$-wave, that there are no mathematical ambiguities. We also provide arguments supporting the absence of ambiguous solutions in more general cases. In \cref{sec:simulations} we present results of numerical simulations, which show that there is indeed a unique solution with the highest likelihood. However, the likelihood function contains many local maxima that may lead to false solutions if appropriate care is not taken when performing fits. The summary and conclusions are given 
 in  \cref{sec:conclusions}. 

\section{Formalism}
\label{sec:formalism}

\begin{figure}[t]
\begin{center}
\includegraphics[width=0.9\linewidth]{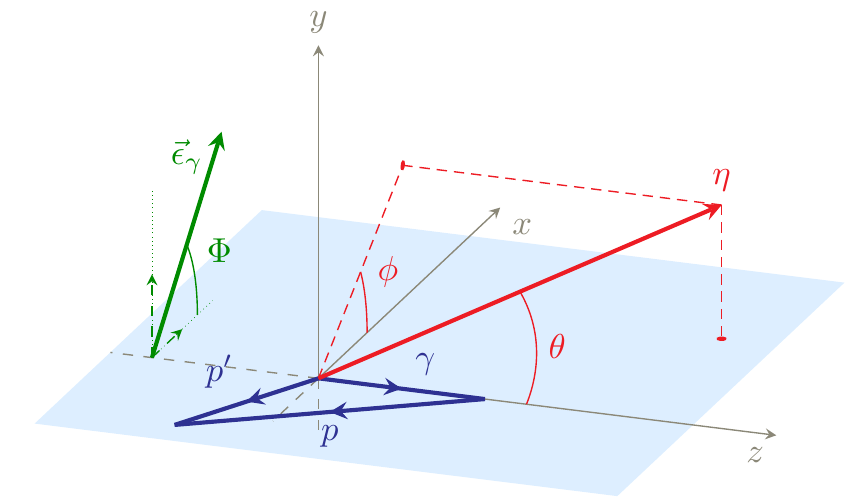}
\end{center}
\caption{\label{fig:frame}Definition of the angles in the Gottfried-Jackson frame.
In the two-meson rest frame, the $z$ axis is given by the photon beam ($\gamma$), and the  $xz$ reaction plane contains also the nucleon target ($p$) and recoiling nucleon ($p'$) momenta. $\theta$ and $\phi$ are the polar and azimuthal angles of the $\eta$. The polarization vector of the photon $(\vec{\epsilon}_\gamma)$ forms an angle $\Phi$ with the reaction plane.
}
\end{figure}

We consider the photoproduction on a nucleon target of a meson resonance decaying into two spinless mesons, \eg $\gamma p \to p\, \eta \, \pi^0$. We follow Ref.~\cite{Mathieu:2019fts}, writing
\begin{align} \nonumber
\label{eq:def_diffI}
I(\Omega,\Phi) &= \frac{\diff\sigma}{\diff t \, \diff m_{\eta\pi^0} \, \diff\Omega \, \diff\Phi} \\
& = \kappa
\sum_{ \substack{\lambda_\gamma \lambda'_\gamma \\ \lambda_1\lambda_2}} A_{\lambda_\gamma; \lambda_1\lambda_2} (\Omega) \rho^\gamma_{\lambda_\gamma\lambda'_\gamma}(\Phi) A_{\lambda'_\gamma; \lambda_1\lambda_2}^* (\Omega),
\end{align}
where $\Omega = (\theta,\phi)$ are the decay angles of the resonance in the Gottfried-Jackson or helicity frame, and $\Phi$ is the polarization angle with respect to the production plane. The spin density matrix is given by $\rho_\gamma(\Phi) = \frac{1}{2}\left(1 -P_\gamma \cos 2\Phi\,\sigma_x - P_\gamma \sin 2\Phi\,\sigma_y\right)$, and $P_\gamma$ indicates the degree of polarization. Since the analysis of ambiguities is performed independently in each bin of $t$ and $\eta\pi^0$ invariant mass, these dependences are understood. The phase space factor $\kappa$ does not depend on angular variables and will be absorbed into the amplitudes. We neglect the dependence on the nucleon spin~\footnote{For the complete discussion including nucleon spin, see~\cite{Mathieu:2019fts}.} and write for the helicity amplitudes
\begin{align}
    \label{eq:helamp}
    A_{\lambda_\gamma}(\Omega)=\sum_{\ell m}[\ell]_{\lambda_\gamma;m} Y_\ell^m(\Omega),
\end{align}
where $[\ell]_{\lambda_\gamma;m}$ refers to the partial wave with angular momentum $\ell$, spin projection $m$ produced with photon helicity $\lambda_\gamma$.

One can construct partial waves with definite reflectivity as linear combinations of the partial waves in such a way that in the high energy limit positive (negative) reflectivity corresponds to natural (unnatural) parity exchanges in the Gottfried-Jackson frame represented in \cref{fig:frame}~\cite{Mathieu:2019fts,Chung:1997qd},

\begin{align}
    \label{eq:reflectivitydef}
    [\ell]^{(\epsilon)}_m &=\frac 1 2\left([\ell]_{+1;m}-\epsilon (-1)^m[\ell]_{-1;-m}\right).
\end{align}
In doing so, we have essentially traded the photon helicity $\lambda_\gamma$ for reflectivity $\epsilon$.
For convenience, we define the amplitudes $U^{(\epsilon)}$ and $\tilde U^{(\epsilon)}$ in the
reflectivity basis:
\bsub\label{eq:def_U}\begin{align}
    U^{(\epsilon)}(\Omega) & = \sum_{\ell m} [\ell]^{(\epsilon)}_m Y^m_\ell(\Omega), \\
    \tilde U^{(\epsilon)}(\Omega) & = \sum_{\ell m} [\ell]^{(\epsilon)}_m \left[Y^m_\ell(\Omega) \right] ^*.
\end{align}\esub

We write the intensity of the final products from \cref{eq:def_diffI},
\begin{align}\label{eq:def_Ipre}
    I(\Omega,\Phi) = I^0(\Omega) &-P_\gamma I^1(\Omega)\cos(2\Phi)\notag \\
    &-P_\gamma I^2(\Omega) \sin(2\Phi),
\end{align}
where $I^0$ is the unpolarized intensity, and $I^{1,2}$ are polarized intensities. The intensities are quadratic in the partial waves and can be expressed in terms of the amplitudes in \cref{eq:def_U}:
\bsub\label{eq:def_I}\begin{align}
  I^0(\Omega)& = \phantom{-2}\sum_\epsilon \left\{|U^{(\epsilon)}(\Omega)|^2 + |\tilde U^{(\epsilon)}(\Omega)|^2 \right\}\, ,
  \\
  I^1(\Omega)& = -2\sum_\epsilon \epsilon \Re\left\{ U^{(\epsilon)}(\Omega)\, \left[\tilde U^{(\epsilon)}(\Omega)\right]^* \right\} \, ,
  \\
  I^2(\Omega)& = -2\sum_\epsilon \epsilon \Im\left\{ U^{(\epsilon)}(\Omega)\, \left[\tilde U^{(\epsilon)}(\Omega)\right]^* \right\} \, .
\end{align}\esub
The dependence on the polar angle $\theta$ can be written explicitly by expanding the intensities in a Fourier series in the azimuthal decay angle $\phi$:
\begin{subequations} \label{eq:defh}
\begin{align}
    I^0(\Omega) &= \phantom{-}\frac{1}{2\pi}\Bigr[& h_0^0(\theta)~&+&h_1^0(\theta)\cos(\phi)& ~+~~ ...~\Bigr],\\
    I^1(\Omega) &= -\frac{1}{2\pi}\Bigr[&h_0^1(\theta)~&+&h_1^1(\theta)\cos(\phi)& ~+~~...~\Bigr],\\
    I^2(\Omega) &= -\frac{1}{2\pi}\Bigr[&0~~~~&+&h_1^2(\theta)\sin(\phi)& ~+~~ ...~\Bigr]. \label{eq:defhc}
\end{align}
\end{subequations}
Here the ellipses denote terms of higher order harmonics in $\phi$.

The functions $h^\alpha_M(\theta)$, which we will refer to as (un)polarized moments, are quadratic in the partial waves and relate them 
 to the measurable angular distribution 
  of the two mesons in their center of mass frame. We note that positive and negative reflectivity contributions sum up incoherently, and one can decompose $h^\alpha_M(\theta)$ into an explicit sum of reflectivity components , \ie $h^\alpha_M(\theta) = \ ^{(+)}h^\alpha_M(\theta) + \ ^{(-)}h^\alpha_M(\theta)$.
The two reflectivities can be distinguished from each other due to the dependence on the polarization angle $\Phi$. We can therefore deal with each reflectivity independently, noting that the mathematical treatment of ambiguities is identical for each.
  
  To pursue our analysis of Barrelet zeros, we need to express the observables $h^\alpha_M(\theta)$ as polynomials of $
  \tan\frac{\theta}{2}$, and then extract their roots. We first employ \cref{eq:def_U,eq:def_I} and rewrite \cref{eq:defh} as:
\begin{subequations}\label{eq:intens}
\begin{align}
    I^0(\Omega)&= \frac{1}{2\pi}\sum_{\epsilon m m'}f_m^{(\epsilon)}(\theta) f_{m'}^{(\epsilon) *}(\theta)\cos[(m-m')\phi],\\
I^1(\Omega) & =\frac{-1}{2\pi}\sum_{\epsilon mm'}\epsilon f^{(\epsilon)}_m(\theta) f^{(\epsilon)*}_{m'}(\theta) \cos [(m+m')\phi], \\
I^2(\Omega) & = \frac{-1}{2\pi}\sum_{\epsilon mm'}\epsilon f^{(\epsilon)}_m(\theta) f^{(\epsilon)*}_{m'}(\theta) \sin [(m+m')\phi],
\end{align}
\end{subequations}
where,
\begin{align} 
    f^{(\epsilon)}_m(\theta) &= \sum_\ell \sqrt{4\pi}\,[\ell]^{(\epsilon)}_m Y^{m}_\ell(\theta,0) \nonumber \\ &= \sum_\ell \sqrt{2\ell+1}\,[\ell]^{(\epsilon)}_md^\ell_{m0}(\theta).\label{eq:def_f}
\end{align}
The Wigner $d$-function,  $d^\ell_{m0}(\theta)$,\footnote{We use the Wigner $d$-function with the convention $d^j_{m'm}(\theta)=\bra{jm'}e^{-i\theta J_y}\ket{jm}$} is 
 a polynomial in $\cos\theta$ only for $m=0$.  For 
  $m \ne 0$ it is a  polynomial of 
  $\cos\theta$ of order $l-|m|$ multiplied by a factor $\sin^{|m|}(\theta)$. We thus represent the $d$-functions in terms of $u = \tan \theta/2$ by~\cite{Chung:1997qd}: 
\begin{align}
d^{\ell}_{m0}(\theta) & = \left(\frac{u}{1+u^2}\right)^{\ell}  (-1)^m\varepsilon^\ell_m(u) \, ,
\end{align}
with the polynomial $\varepsilon^\ell_m(u)$ defined as:
\begin{align}
    \varepsilon_m^\ell(u) & =   \sum_k (-1)^{k} \frac{  u^{2k+m-\ell} \ell ! [(\ell-m)!(\ell+m)!]^{1/2}}{(\ell-m-k)!(\ell-k)!(m+k)! k!} \, .
\end{align}
The summation over $k$ is restricted to the range $k\in [\text{max}(0, -m), \text{min}(\ell,\ell-m)]$.

By matching \cref{eq:defh,eq:intens}, we obtain a relation between the observable quantities and the reflectivity partial waves:
\bsub \label{eq:h_with_f}
\begin{align}
\ ^{(\epsilon)}h^0_{M} & =   \sum_{mm'} f_m^{(\epsilon)}f_{m'}^{(\epsilon)*}\delta_{M,|m-m'|} \, ,\\
\ ^{(\epsilon)}h^1_{M} & =  \epsilon\sum_{mm'} f_m^{(\epsilon)}f_{m'}^{(\epsilon)*} \delta_{M,|m+m'|} \, ,
\\
\ ^{(\epsilon)}h^2_{M} & = \epsilon \sum_{mm'} f_m^{(\epsilon)}f_{m'}^{(\epsilon)*} \delta_{M,|m+m'|} \sign(m+m') \, .
\end{align}\esub
Since each $f^{(\epsilon)}_m(\theta)$ is a complex function, and each $h^\alpha_M(\theta)$ is a real observable expressible as a sum of products of $f$-functions, one may simplify the problem by expressing $h^\alpha_M(\theta)$ as a sum of squares of complex functions,
\begin{align}
    \label{eq:hdiagwithgs}
    h^\alpha_M(u) = \sum_i|g_i(u)|^2.
\end{align}
Here, each $g_i(u)$ is a linear combination of the $f_m^{(\epsilon)}(\theta)$, and therefore is also a rational function in $u$. Hence, conjugation of the roots of each $g_i(u)$ may generate ambiguities of the partial waves. We note that it is most convenient to express every moment in terms of a single basis set of $g$'s.
\cref{eq:h_with_f} represent bilinear matrix equations which connect the coefficients of the intensity \cref{eq:defh} to the partial wave amplitudes, while \cref{eq:hdiagwithgs} represents a diagonalization of the same equations. Since the moments can be extracted directly from experimental data, the presence of mathematical ambiguities is determined by whether or not replacing roots of the basis functions $g_i$ with their conjugates provide alternate solutions to these matrix equations.
 To address this we will consider a few examples explicitly to show that for a few sets of partial waves $\{[\ell]^{(\epsilon)}_m\}$, the relations \cref{eq:h_with_f} are uniquely determined and no ambiguities exist. In other words there is no way to construct a different set, $\{[\tilde{\ell}]^{(\epsilon)}_m\}$ which will yield the same moments.
 
\section{Case studies}
\label{sec:SDexample}

Ambiguities in partial wave analysis with a high energy pion beam were studied in Ref.~\cite{Chung:1997qd}, where several wave sets with different combinations of waves up to the $G$ wave were considered. In all cases the spin projections were limited to $m=0,1$.\footnote{For pion beams, $m\ge 0$ in the reflectivity basis. This does not hold for photon beams.} With this restriction, the intensity only includes three terms in the azimuthal expansion of
 \cref{eq:defh}. Since for a pion beam there is no $S$ wave with positive reflectivity, there is one relevant $g$-function for the positive reflectivity components, and two for the negative reflectivity components. The polynomials which generate ambiguities in the negative reflectivity components are not independent, so the ambiguities for the waves in each reflectivity component are obtained using the roots of a single polynomial. That is, there are ambiguous solutions in the partial wave extraction because the observable depends on only two independent polynomials, one for each reflectivity component, and transformations built from combinations of conjugations of roots of each polynomial produce the same intensity profile.

In this section, we will argue that there are no ambiguities in the extraction of partial waves from an experiment using a linearly polarized photon beam. First, we note that any possible ambiguities arising from switching contributions between the two different reflectivity waves may be resolved by making use of the $\Phi$ dependence of the linearly polarized photon beam. We will thus only consider one reflectivity component and suppress all reflectivity superscripts for convenience.

We will consider first the simplest non-trivial case by including only the waves $\{ S_0, D_0,D_1 \}$. This case is analogous to Ref.~\cite{Chung:1997qd}, however, as we will see, the polarized intensity allows us to determine the partial waves without ambiguity.

We then will consider the wave set $\{ S_0, D_{-1}, D_0,D_1 \}$. These $D$ waves dominate the production of the $a_2(1320)$ resonance in the $\eta\pi$ final state {\it via} pion exchange~\cite{Mathieu:2020zpm}. We will not find any ambiguous solution for the extraction of this wave set, once the polarized moments are taken into account. The $a_2(1320)$ is also produced by vector exchanges. In this case, the dominant $D$ waves are $\{D_0,D_1, D_{2} \}$~\cite{Mathieu:2020zpm}. We have confirmed that this wave set is also free of ambiguities, although we omit the calculation for brevity.

Our key result is that there are at least two unique $g$'s which appear in the Fourier series of the polar angle when two or more spin projections are allowed. These polynomials are independent and have distinct roots. Consequently, these Fourier moments are enough to uniquely determine the partial waves. No transformations on the partial waves leave every observable invariant, and the observables uniquely define the partial waves for linearly polarized meson photoproduction.
We illustrate this fact only with $S$ and $D$ waves, but the addition of other waves should not change our results. Adding more waves increases the number of roots of each $g$-function, and hence the number of possible ambiguities, but in general we argue that there is no relation between the roots, and therefore partial waves can be unambiguously extracted from the polarized observables.

\subsection{\boldmath $S$ and $D$ waves with $m=0,1$} \label{sec:SD01}
We start by analyzing the wave set with $S$ and $D$ waves with $m$ projections $0,1$ and positive reflectivities, as this set has been analyzed explicitly for a pion-beam production process~\cite{Chung:1997qd}. 
Suppose that we have obtained one set of partial waves, $\{S_0, D_0, D_1\}$, from an experiment. We can then attempt to generate an ambiguous set of partial waves, $\left\{\tilde{S}_0,\tilde{D}_0, \tilde{D}_{1}\right\}$, from the original set. 
We start by writing the $f$'s from \cref{eq:def_f}:
\begin{subequations}
\begin{align}
   f_0(u) &=\frac{\sqrt{5} \left(u^4-4 u^2+1\right) D_0}{\left(u^2+1\right)^2}+S_0 \, ,\\
   f_1(u) &=\frac{\sqrt{30}\, u\left( u^2- 1\right) D_1}{\left(u^2+1\right)^2} \, .
\end{align}
\end{subequations}

With this wave set, there are seven non-zero functions $h^\alpha_M(\theta)$, though they are not all linearly independent. When the wave set includes only positive $m$-projections, there is a simple relation between the polarized moments $h^2_M = h^1_M$ for $M>0$ ~\cite{Mathieu:2019fts}. (For $M=0$, one has $h^2_0=0$, see Eq.~\eqref{eq:defhc}). In addition, we find the relation $h^1_1 = h^0_1$ and $h^1_2 = h^0_0 - h^1_0$ for this particular wave set. So we are left with three linearly independent $h^\alpha_M(\theta)$.
We rewrite the conditions relating the $h$'s to the $f$'s in matrix form:
\begin{align}
    h_M^0(\theta) &= F^\dagger H_M^0 F, &
    h_M^1(\theta) &= F^\dagger H_M^1 F \, .
\end{align}
Where $F=(f_0,f_1)^T$.  The three matrices are:
\begin{align}
H_0^0 & =\begin{pmatrix} 1&0 \\ 0 & 1 \end{pmatrix},  &
H_1^0 & = \begin{pmatrix}  0 &1 \\  1 & 0 \end{pmatrix},&
H_0^1 & = \begin{pmatrix} 1& 0 \\ 0 & 0 \end{pmatrix} \, .
\end{align}
Since the matrices $H^0_0$ and $H^0_1$ commute, we can simultaneously diagonalize them and simplify the unpolarized moments, obtaining:
\bsub\begin{align}
    g_0(u) &\equiv \frac{1}{\sqrt 2}\left[f_1(u)+f_0(u)\right] \, ,\\
    g_1(u) &\equiv \frac{1}{\sqrt 2}\left[f_1(u)-f_0(u)\right] \, .
\end{align}\esub
Since $f_0(u)$ is even and $f_1(u)$ is odd, the new functions fulfill $g_1(-u) = - g_0(u)$. Thus, their roots and ambiguities from complex conjugation of the roots are the same.  
The three independent moments read:
\bsub
\begin{align}
h_0^0 =& |g_0|^2+|g_1|^2, \\
h_1^0=&|g_0|^2-|g_1|^2,\\
h_0^1=& \frac{1}{2} |g_{0}-g_1|^2.
\end{align}
\esub
We note that the moments $h^\alpha_M$ will simply change by a sign $(-1)^M$ under the substitution $g_0 \to g_1$.
It is necessary and sufficient to require that any prospective ambiguity transformation leaves invariant $|g_0|^2$ and $|g_0-g_1|^2$ independently.
In terms of the partial waves, these functions can be written:
\bsub\begin{align} \nonumber
g_0=&\sqrt{\frac{5}{2}}\frac{1}{(u^2+1)^2}\Bigr[D_0(u^4-4u^2+1)\\&+\sqrt{6}D_1(u^3-u)\Bigr]+\frac{1}{\sqrt{2}}S_0,\\ 
g_{0}-g_1=&\sqrt{10}\frac{(u^4-4u^2+1)D_0}{(u^2+1)^2}+\sqrt{2}S_0.
\end{align}\esub
Which can be simplified defining $v=u-1/u=-2\cot\theta$:
\bsub\begin{align}
    g_0=&\sqrt{\frac{5}{2}}\frac{1}{v^2+4}\Bigr[Av^2+\sqrt{6}D_1v-2B\Bigr], \label{eq:1stpolynomial} \\
    g_0-g_1=&\frac{\sqrt{10}}{v^2+4}\Bigr[Av^2-2B\Bigr], \label{eq:2ndpolynomial}
\end{align}\esub
where $A=D_0+S_0/\sqrt{5}$ and $B=D_0-2 S_0/\sqrt{5}$.

Recalling that the ambiguous waves should be generated by conjugating roots of these polynomials, we start by considering the first polynomial in \cref{eq:1stpolynomial}, and factorize it into its Barrelet zeros $r_{1,2}$:
\begin{align}
g_0 \propto &~(v-r_1)(v-r_2),
\end{align}
where we have dropped the irrelevant factors.
The roots read:
\begin{align} \label{eq:roots}
    r_{1,2} & = \frac{-\sqrt{3} D_1 \pm \sqrt{4 A B + 3 D_1^2}}{\sqrt{2} A}.
\end{align}
In this case, there are only two Barrelet zeros and there is thus only one non-trivial independent solution given by the substitution of one root by its complex conjugate. We invert \cref{eq:1stpolynomial} and replace $r_1$ with its conjugate to obtain:
\bsub\label{eq:ambig_1}
\begin{align}
    \tilde S_0 &= \sqrt{5}\frac{A}{6}(2+r_1^*r_2), \\
    \tilde D_0 &= \frac{A}{6}(4-r_1^*r_2), \\
    \tilde D_1 &= -\frac{A}{\sqrt{6}}(r_1^*+r_2).
\end{align}\esub
We note that the new waves obtained by the complex conjugation of $r_1$ and $r_2$ simultaneously lead to the set $\{S_0^*, D_0^*, D_1^*\}$, the complex conjugate of the original wave set. 
For a given wave set $\{S_0,D_0,D_1\}$, the set in \cref{eq:ambig_1} produces the same unpolarized moments $h^0_{0,1}(\theta)$. 
In the absence of information on the polarized moments, the above wave set would constitute an ambiguous solution.

In this example, the use of observables only accessible via a polarized beam are essential to ensure that no mathematical ambiguities can occur.
In particular, we must consider the constraints implied by the polarized moment $h_0^1=\frac{1}{2}|g_{0}-g_1|^2$. The combination $g_0-g_1$ only has one Barrelet zero, \ie $g_0-g_1 \propto (v-r_3)(v-r_3^*)$, where $ r_3 = \sqrt{2} B/A$. This is independent of $r_{1,2}$, and the only transformation that leaves $h^1_0(\theta)$ invariant is the one that replaces each wave by its complex conjugate, since all the waves are defined up to a global phase. Therefore, there is no nontrivial transformation of the partial waves which leaves both the unpolarized moments $h_{0,1}^0(\theta)$ and the polarized moment $h^1_0(\theta)$ invariant, and thus there are no ambiguous solutions for this wave set.

\begin{figure*}
     \centering
    \includegraphics[width=\textwidth]{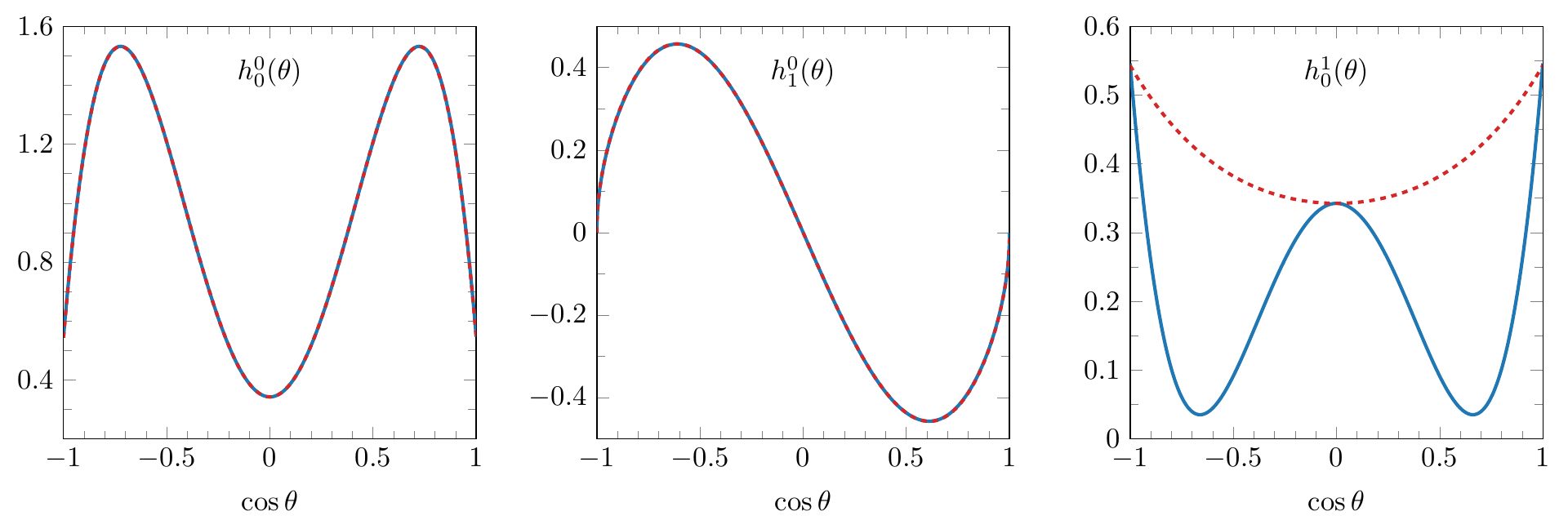}
     \caption{{\it Solid blue lines}, moments obtained from the original waves of \cref{tab:ambig1}; {\it dotted red lines}, moments obtained from the ambiguous solution of \cref{tab:ambig1}. The polarized moment $h^1_0$ breaks the ambiguity between the two solutions.}
     \label{fig:ambig1}
\end{figure*}

We illustrate this case for one single energy bin by choosing three random complex numbers for the original waves $\{S_0, D_0, D_1\}$,\footnote{We choose $S_0$ to be real positive without loss of generality and rotate the ambiguous solution to bring  $\tilde S_0$ also to the positive real axis, \ie its phase is zero.} compute the associated ambiguous solutions $\{\tilde S_0, \tilde D_0, \tilde D_1\}$ and display the three moments in \cref{fig:ambig1}. The numerical values of the waves are specified in \cref{tab:ambig1}. Here again, we see the value of incorporating polarized observables. While the two wave sets produce degenerate solutions for the two unpolarized moments, the incorporation of the polarized moment $h_0^1$ breaks the degeneracy.

The inclusion of more waves with only the projections $m=0,1$ will not change our results. Adding more waves with different $m$ projections could potentially produce ambiguous solutions, each of which leave invariant one single moment $h^\alpha_M(\theta)$, but it would also generate additional nonzero $h^\alpha_M(\theta)$ which must remain invariant under each of the ambiguity transformations. One can try to generate other prospective ambiguities, but each potentially ambiguous wave set will be subject to an increasing number of constraints. Hence, we argue that, for most sensible wave sets, the intersection between all these sets of potentially ambiguous waves will be empty. 

\begin{table}[h!] 
\caption{Numerical values of our example wave set and the potentially ambiguous wave set generated by the unpolarized moments.\label{tab:ambig1}}
\begin{ruledtabular}
\begin{tabular}{c|c  c }
    $[\ell]_{m}$ & original & potentially ambiguous\\\hline
   $S_0$ & $0.229$ & $0.630$\\
   $D_0$ & $-0.217 + 0.310i$ & $0.043 + 0.056i$\\
   $D_1$ & $\hphantom{-}0.770 + 0.448i$ & $0.280 - 0.713i$\\   
\end{tabular}
\end{ruledtabular}
\end{table}

\subsection{\boldmath $S$  and $D$ waves with $m=-1,0,1$} \label{sec:SDm101}
We now consider the previous example in \cref{sec:SD01} with the addition of the $m=-1$ projection. The presence of three different $m$ projections raises the number of independent $\cos M \phi$ moments to three ($M = 0,1,2$ in this case), each 
of them being a function of the polar angle. As we will see, it is  impossible to find an ambiguous set leaving all the polar angle distributions simultaneously invariant. We only consider here the $m=-1,0,1$ components for the $D$ wave but our conclusions can be generalized to any wave set with three (or more) spin projections. It was already noticed by the COMPASS collaboration that no ambiguities are found in the $\eta\pi$ system once the $m=2$ component is included in the partial wave analysis~\cite{COMPASS:2014vkj}.

We again start with a set of partial waves, $\{S_0,D_0, D_1,D_{-1}\}$, and attempt to generate an ambiguous set $\{\tilde{S}_0,\tilde{D}_0, \tilde{D}_{1},\tilde{D}_{-1}\}$. 
The $f$'s are:
\begin{subequations}
\begin{align}
   f_0(u) &= \sqrt{5} \frac{\left(u^4-4 u^2+1\right) }{\left(u^2+1\right)^2} D_0+S_0 \, , \\
   f_{\pm 1}(u) &= \mp\sqrt{30}\frac{ u\left( 1- u^2\right) }{\left(u^2+1\right)^2} D_{\pm 1} \, .
\end{align}
\end{subequations}
Our wave set for this example contains all $|m|\leq 1$ but only positive reflectivity components. The structure of the moments in~\cref{eq:h_with_f} tells us that, when only one reflectivity component is included but all $m$ projections are allowed, the polarized moments $h^1_M$ are not independent of the unpolarized moments $h^0_M$. It suffices to study the ambiguities which leave only $h^0_M$ and $h^2_M$ invariant. 

Let us first investigate only the unpolarized moments. 
We rewrite the conditions relating the $h$'s to the $f$'s in matrix form:
\begin{align} \label{eq:MatrixForm2}
    h_M^0(\theta) = F^\dagger H_M^0 F \, ,
\end{align}
where, $F=(f_{-1},f_0,f_1)^T$ and 
\begin{align}
H_0^0 & =\begin{pmatrix} 1&0  &0\\ 0 & 1 &0 \\ 0 & 0 & 1 \end{pmatrix},&
H_1^0 &= \begin{pmatrix} 0& 1 & 0 \\ 1 & 0 & 1 \\ 0& 1& 0\end{pmatrix},&
H_2^0 &= \begin{pmatrix} 0&0  &1\\ 0 & 0 &0 \\ 1 & 0 & 0 \end{pmatrix} .
\end{align}
Notice here that $H_0^0$, $H_1^0$ and $H_2^0$ are all \emph{not} simultaneously diagonalizable.  Nevertheless, as before, we diagonalize $H^0_1$, defining $g_0$, $g_1$, and $g_{-1}$ as 
\bsub\begin{align}
    g_{\pm 1}(u) &\equiv \frac{1}{ 2}\left[f_1(u)\pm \sqrt{2}f_0(u) + f_{-1}(u)\right] \, ,\\
    g_0(u) &\equiv \frac{-1}{\sqrt 2}\left[f_1(u)-f_{-1}(u)\right]\, .
\end{align}\esub
Again, the parity of the $f$'s functions indicates that $g_{\pm 1}(-u) = g_{\mp 1}(u)$ and $g_0(-u) = -g_0(u)$. The two functions $g_{1}$ and $g_{-1}$ possess the same Barrelet zeros and therefore the same potential ambiguities

As in the previous example, the moments are even functions of the polar angles and read, in the $g$ basis,
\begin{subequations}
\begin{align}
h_0^0 =& \left(|g_{-1}|^2+|g_0|^2+|g_1|^2\right) \, , \\
h_1^0=&\sqrt{2}\left(|g_1|^2-|g_{-1}|^2\right)\, , \\
h_2^0=&\frac 1 2 |g_{-1}+g_1|^2-|g_0|^2 \, .
\end{align}
\end{subequations}
Again, any transformation on the partial waves which leaves each term above independently unchanged will produce a mathematically ambiguous set of waves.
Introducing the change of variables $v = u - 1/u$ as before, the relevant rational fractions are
\bsub\begin{align}
    g_{\pm 1} & =  \pm\sqrt{\frac{5}{2}} \frac{1}{v^2+4} \left[ A v^2 \pm \sqrt{6} v D^- -2 B \right],\\
    g_0 & = - \sqrt{30} \frac{v}{v^2+4} D^+,
\end{align}\esub
where $A,B$ are defined as in the previous subsection and $D^\pm  = (D_1 \pm D_{-1})/\sqrt{2}$. With these definitions, the roots of $g_{\pm 1}(v)$ are given by \cref{eq:roots} with the substitution $D_1\to \pm D^-$. 

As already noted, the same ambiguous solution will simultaneously leave invariant $|g_1|^2$ and $|g_{-1}|^2$. The new wave set $\{\tilde S_0, \tilde D_0, \tilde D^-\}$ is easily obtained from \cref{eq:ambig_1} with the substitution $D_1\to  D^-$. There is, in addition, a continuous transformation $D^+\to \exp(i \alpha^+) D^+$ leaving $|g_0|^2$ invariant. Since this transformation is independent from the set $\{\tilde S_0, \tilde D_0, \tilde D^-\}$, we have, so far, found an ambiguous solution, parametrized with a continuous parameter, leaving the moments $h^0_0$ and $h^0_1$ invariant.
However, the invariance of $h^0_2$ requires a continuous transformation of the type $D^-\to \exp(i \alpha^-) D^-$, which contradicts the ambiguous solution $\{\tilde S_0, \tilde D_0, \tilde D^-\}$. Therefore the unpolarized moments $h^0_{0,1,2}$ are left invariant only by the 1-parameter continuous transformation
\begin{align}\label{eq:1param}
    \{S_0, D_0, D^-,D^+\} \rightarrow  \{S_0, D_0, D^-, e^{i\alpha^+}D^+\}.
\end{align}

Since the polarized moments $h^1_{0,1,2}$ are related to the unpolarized ones, we only need to consider the moments $h^2_{1,2}$.  Their respective matrices, in the form analogous to \cref{eq:MatrixForm2}, are 
\begin{align}
H_1^2 & =\begin{pmatrix} 0&-1  &0\\ -1 & 0 &1 \\ 0 & 1 & 0 \end{pmatrix},&
H_2^2 &= \begin{pmatrix} -1& 0 & 0 \\ 0 & 0 & 0 \\ 0& 0& 1\end{pmatrix}.
\end{align}
Their expressions in the $g$'s basis are 
\bsub\label{eq:h2}\begin{align}
    h^2_1 & = 2\Re\left[(g_{1}-g_{-1})g_0^*\right]\, ,\\
   h^2_2 & = -\sqrt{2}\Re\left[(g_{1}+g_{-1})g_0^*\right]\, .
\end{align}\esub
The continuous transformation in \cref{eq:1param} changes the phase of $g_0$ and does not leave the polarized moments \cref{eq:h2} invariant. 

We thus conclude that there is no ambiguity associated with the extraction of partial waves with a linearly polarized beam for this wave set, other than the trivial ambiguities given by the rotation of all waves by a common phase, or by the complex conjugation of all waves. 

\section{Simulations}
\label{sec:simulations}
While in the previous sections we have provided arguments that no mathematical ambiguities exist in partial-wave analysis of two
mesons produced with a linearly polarized photon beam, the complicated multidimensional shape of likelihood functions or other functions used for fitting can present themselves as false solutions, which one might naively label as mathematically ambiguous. In this section,
we present some Monte Carlo studies showing this effect. We wish to emphasize that here we only investigate the dependence on statistics of a perfect model. Other factors such as acceptance corrections, resolutions, and other systematic effects are experiment-dependent and may qualitatively alter the results. Studies based on pseudodata or studies involving full experiment simulations will be an important part of subsequent analyses, and might be employed to help discard false solutions or assess the impact of limited statistics.

First, pseudodata was generated following the angular intensity given by \cref{eq:intens,eq:def_f}. We used the wave set from \cref{sec:SDm101}, and generated the pseudodata using the fixed ``true solution" wave set, with non-zero, positive reflectivity partial waves shown in \cref{tab:sim1} and a mean linear polarization degree of $P_\gamma = 0.85$.

\begin{table}[h!] 
\caption{Numerical values of our ``true'' wave set for the simulation studies.\label{tab:sim1}}

\begin{ruledtabular}

\begin{tabular}{c|cc}
    $[\ell]_{m}$& Magnitude & Phase \\\hline
    $S_{0\phantom{-}}$       &      0.499     & $\phantom{-}0\degree$      \\
    $D_{-1}$    &  0.201     & $\phantom{-}15.4\degree$  \\
    $D_{0\phantom{-}}$       & 0.567     & $\phantom{-}174\degree$   \\
    $D_{1\phantom{-}}$       & 0.624     & $-81.6\degree$ 
\end{tabular}
\end{ruledtabular}
\end{table}

\begin{figure}
    \centering
    \begin{tabular}{c}
    \includegraphics[width=0.9\columnwidth]{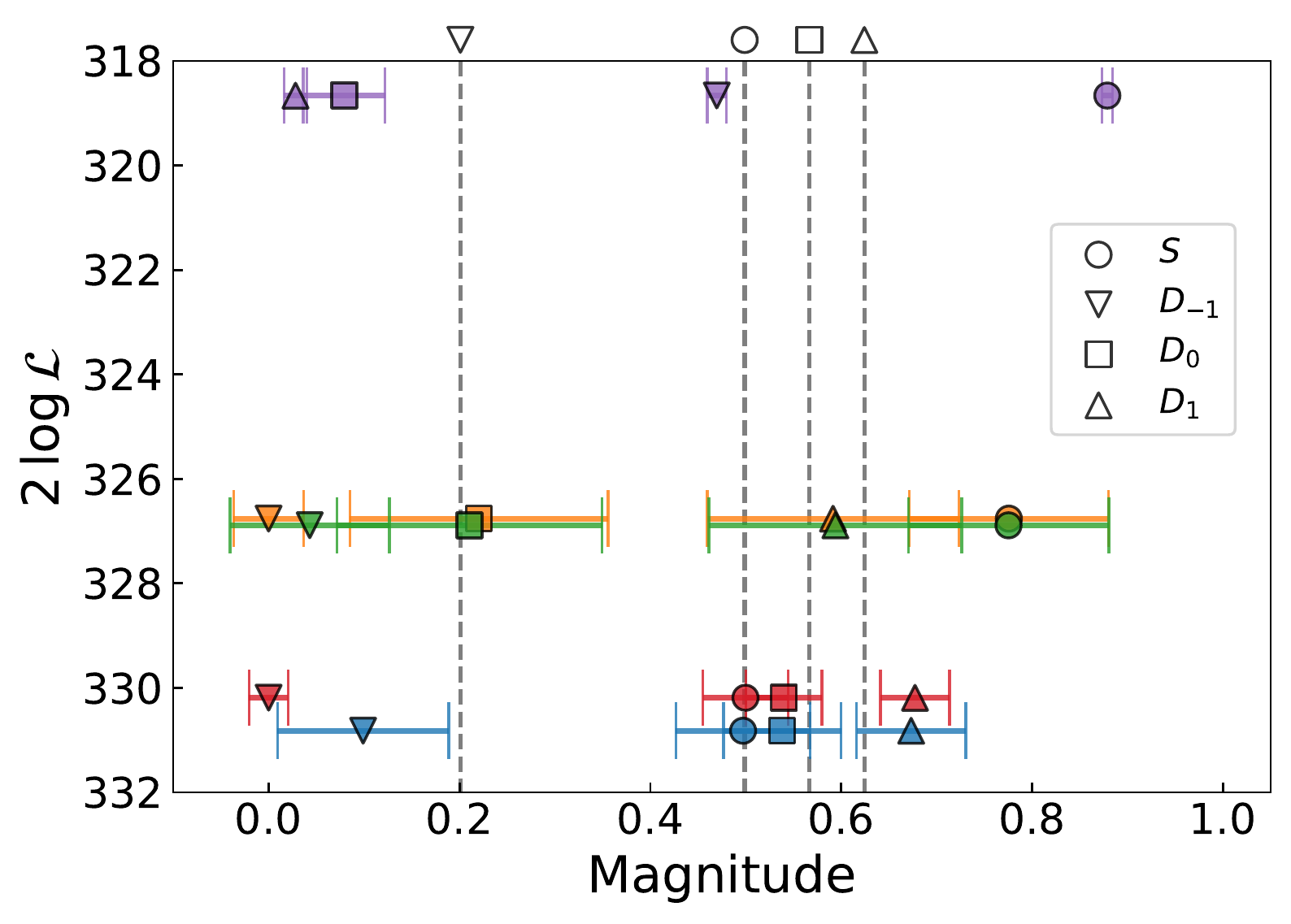} \\
    \includegraphics[width=0.9\columnwidth]{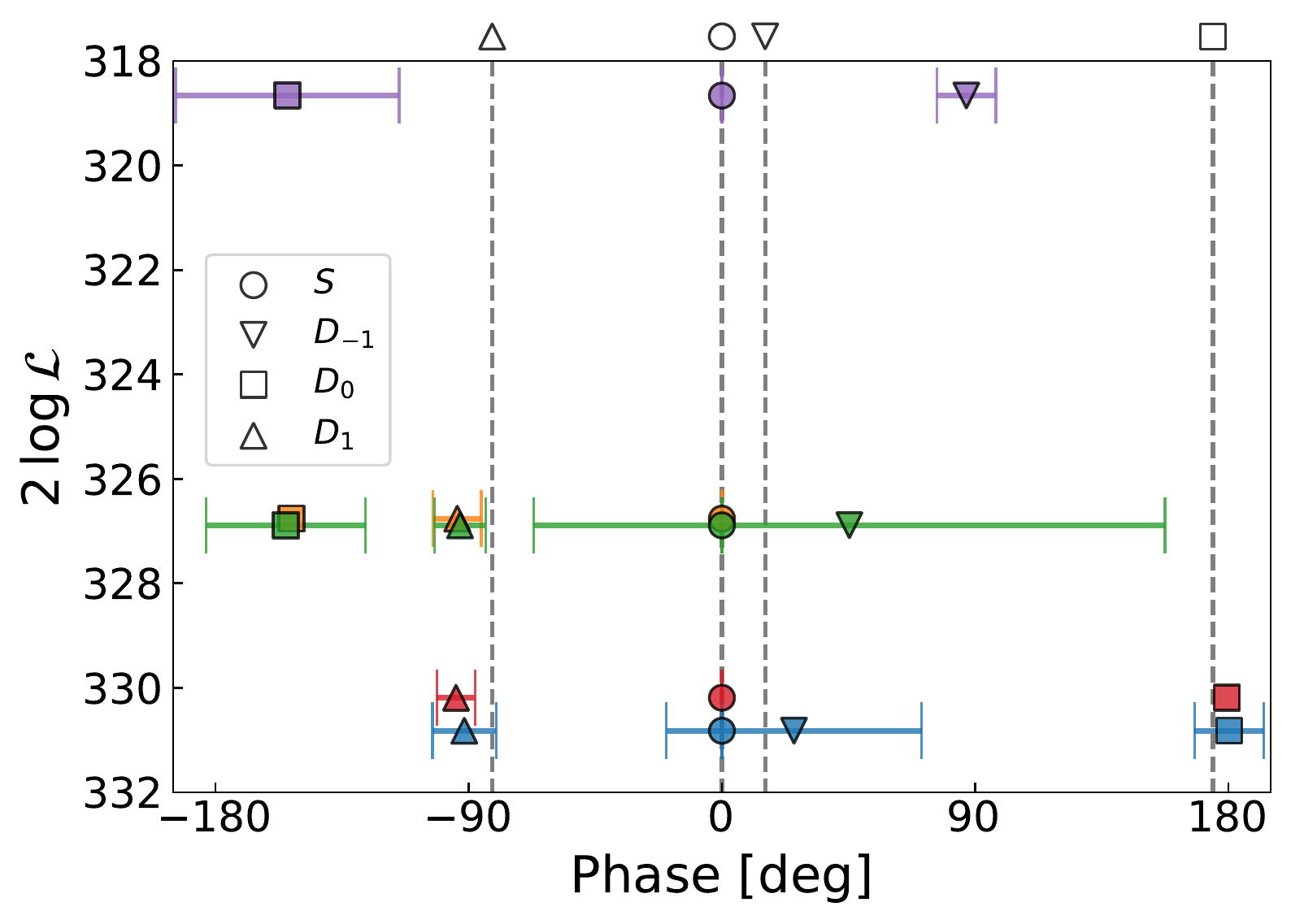} 
    \end{tabular}
    \caption{\label{fig:sim_N100waves} 
    Results of the 5 best (highest likelihood) fits from 50, to 100 events generated with the partial waves given in \cref{tab:sim1} showing the likelihood versus the amplitude magnitude (upper) and phase (lower), the dashed lines show the true values. The wave is indicated by the marker shape (see legend) while the color represents different solutions. The highest likelihood is at the bottom of the plots.
    The phase for the $D_{-1}$ ($D_1$) wave is not shown for the red and orange (purple) fits, as associated magnitude is zero and, hence, the phase is undetermined.
   Fits and uncertainties are computed using the \texttt{HESSE} option of \texttt{MINUIT}~\cite{James:1975dr}.}
\end{figure}

\begin{figure}
    \centering
    \begin{tabular}{c}
    \includegraphics[width=0.9\columnwidth]{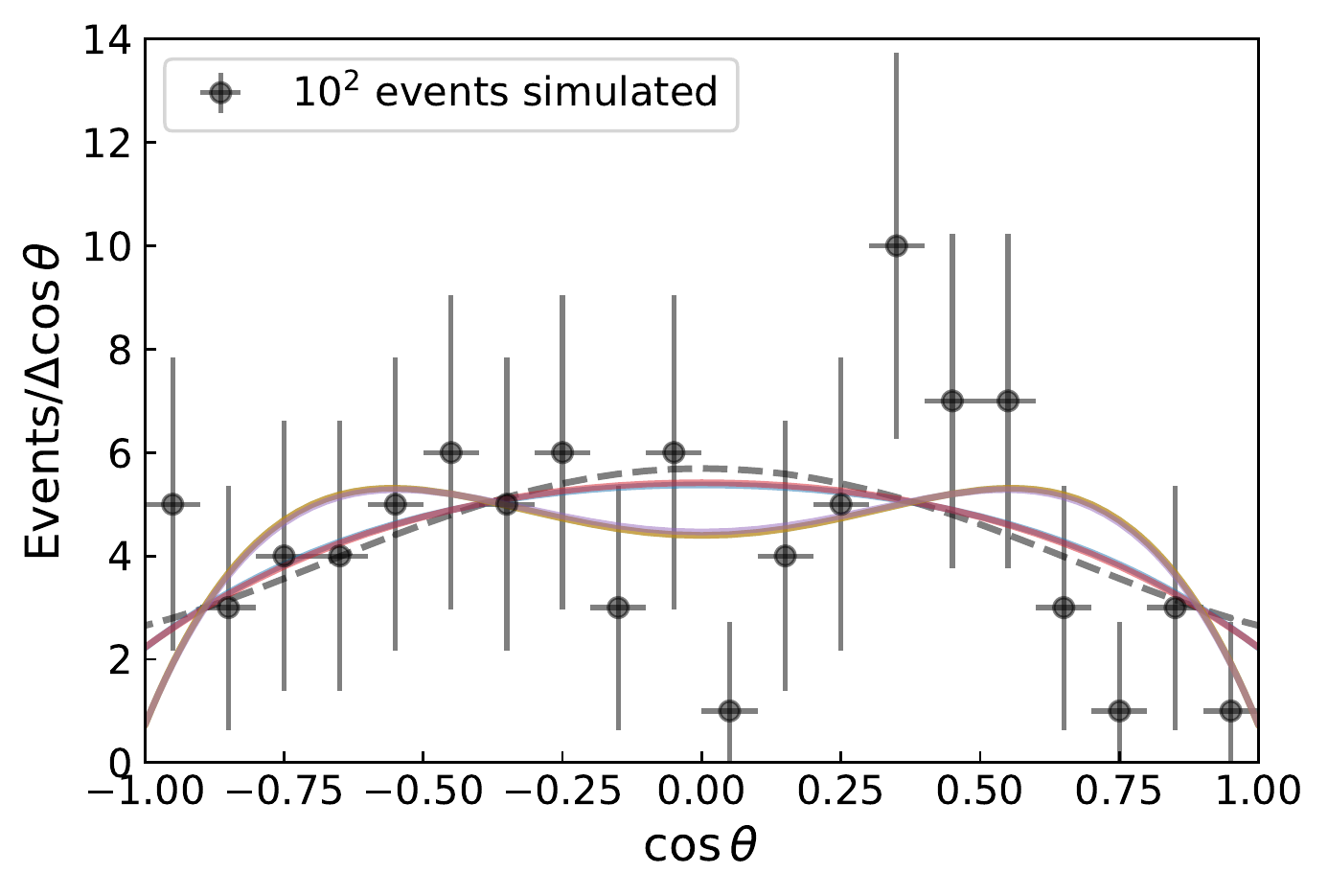} \\
    \includegraphics[width=0.9\columnwidth]{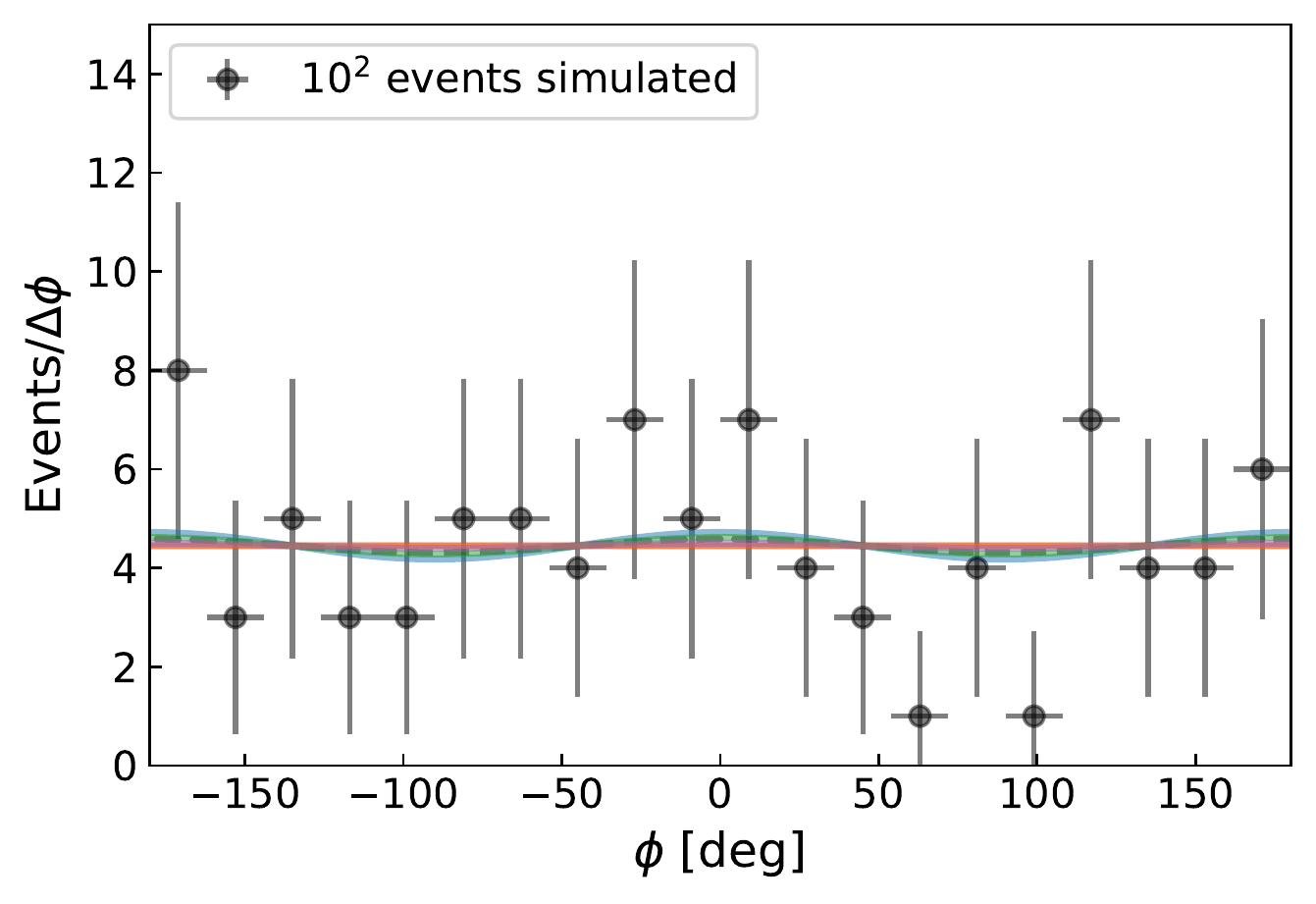} \\
    \includegraphics[width=0.9\columnwidth]{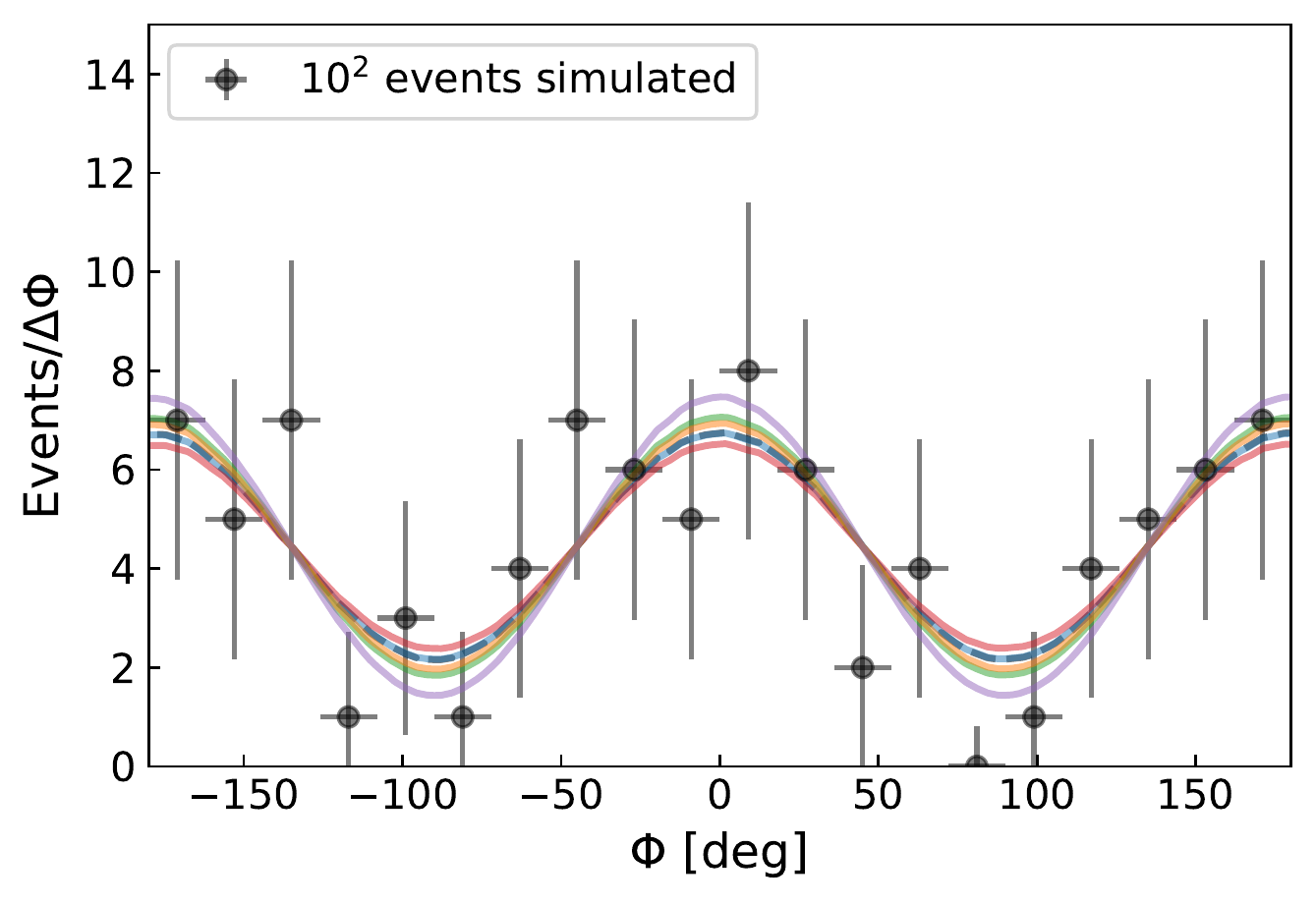}
    \end{tabular}
    \caption{\label{fig:sim_N100plots} 
        Projections of the angular distributions (upper: $\cos\theta$, center: $\phi$, lower: $\Phi$) as defined in \cref{eq:def_Ipre,eq:intens}. 
        Shown are the data (black circles), the true solution (dashed black), and the different solutions (colored lines), with colors matching the plots in \cref{fig:sim_N100waves}. Bin widths are $\Delta\cos\theta=0.1$ and $\Delta\phi=\Delta\Phi=18^\text{o}$.}
\end{figure}

\begin{figure}
    \centering
    \begin{tabular}{c}
    \includegraphics[width=0.9\columnwidth]{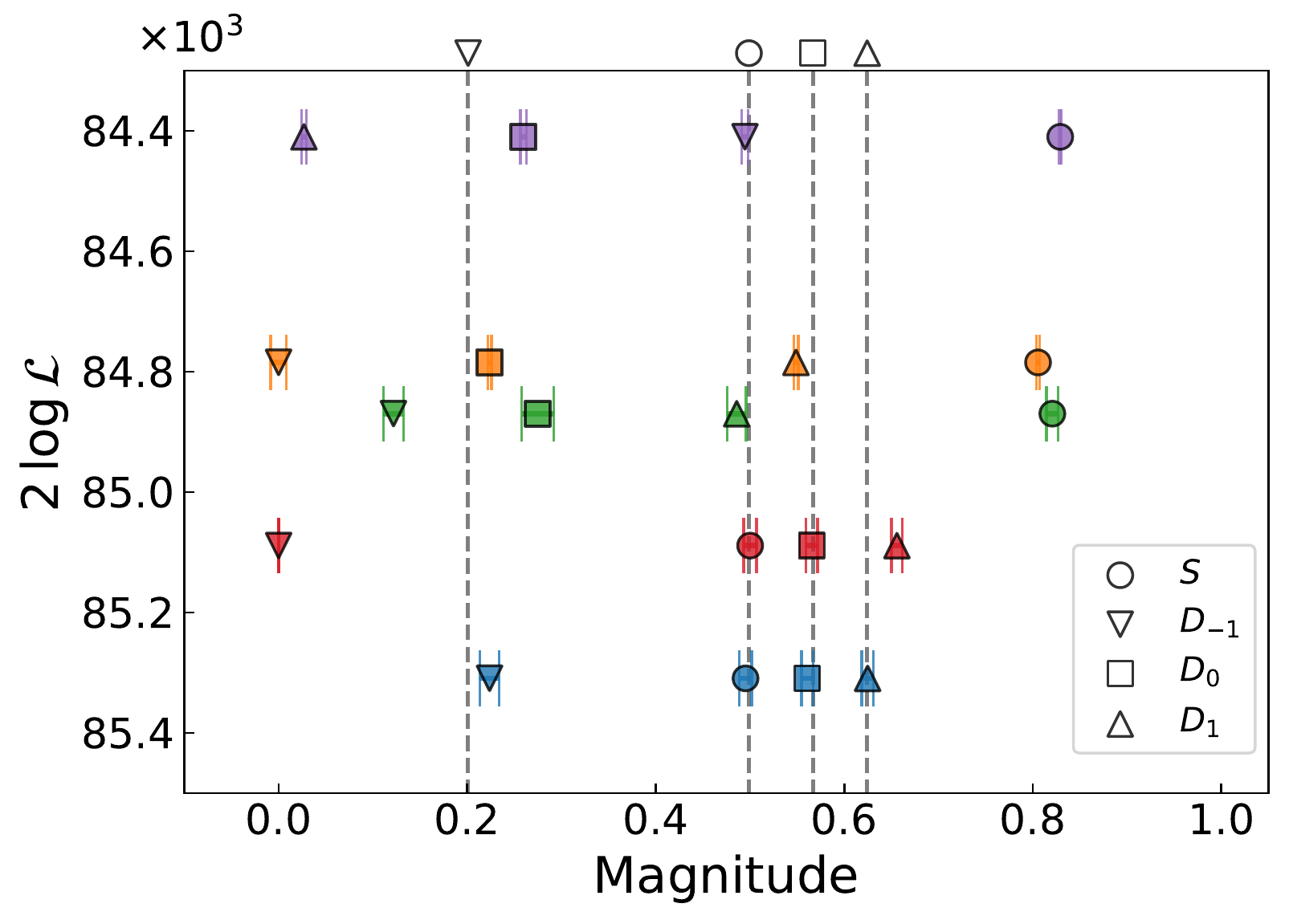} \\ 
    \includegraphics[width=0.9\columnwidth]{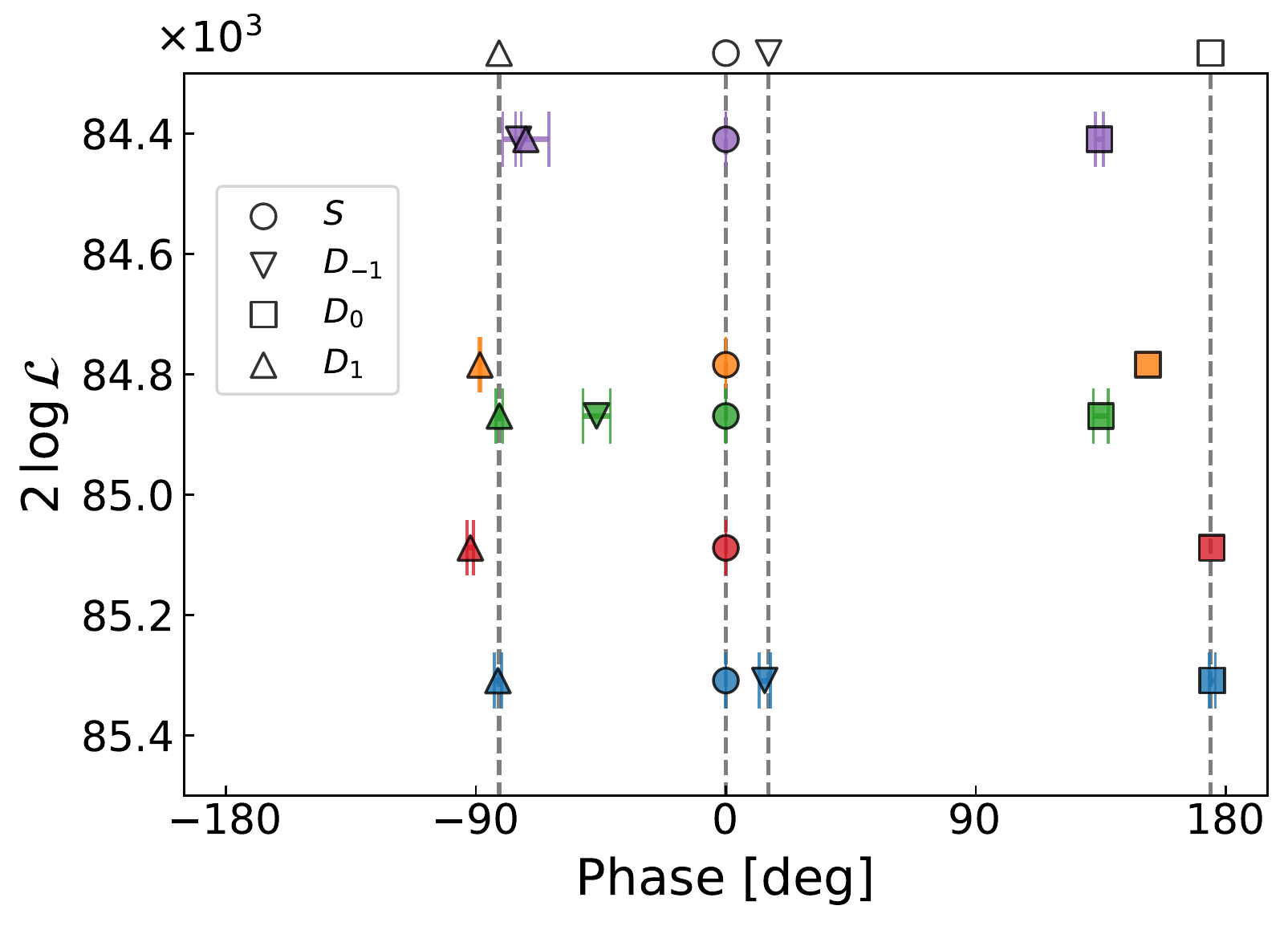} 
    \end{tabular}
    \caption{\label{fig:sim_N10kwaves} As \cref{fig:sim_N100waves} for fits to $10^4$ events. The phase for the $D_{-1}$ wave is not shown for the red and orange fits, as associated magnitude is zero and, hence, the phase is undetermined.}
\end{figure}

\begin{figure}
    \centering
    \begin{tabular}{c}
    \includegraphics[width=0.9\columnwidth]{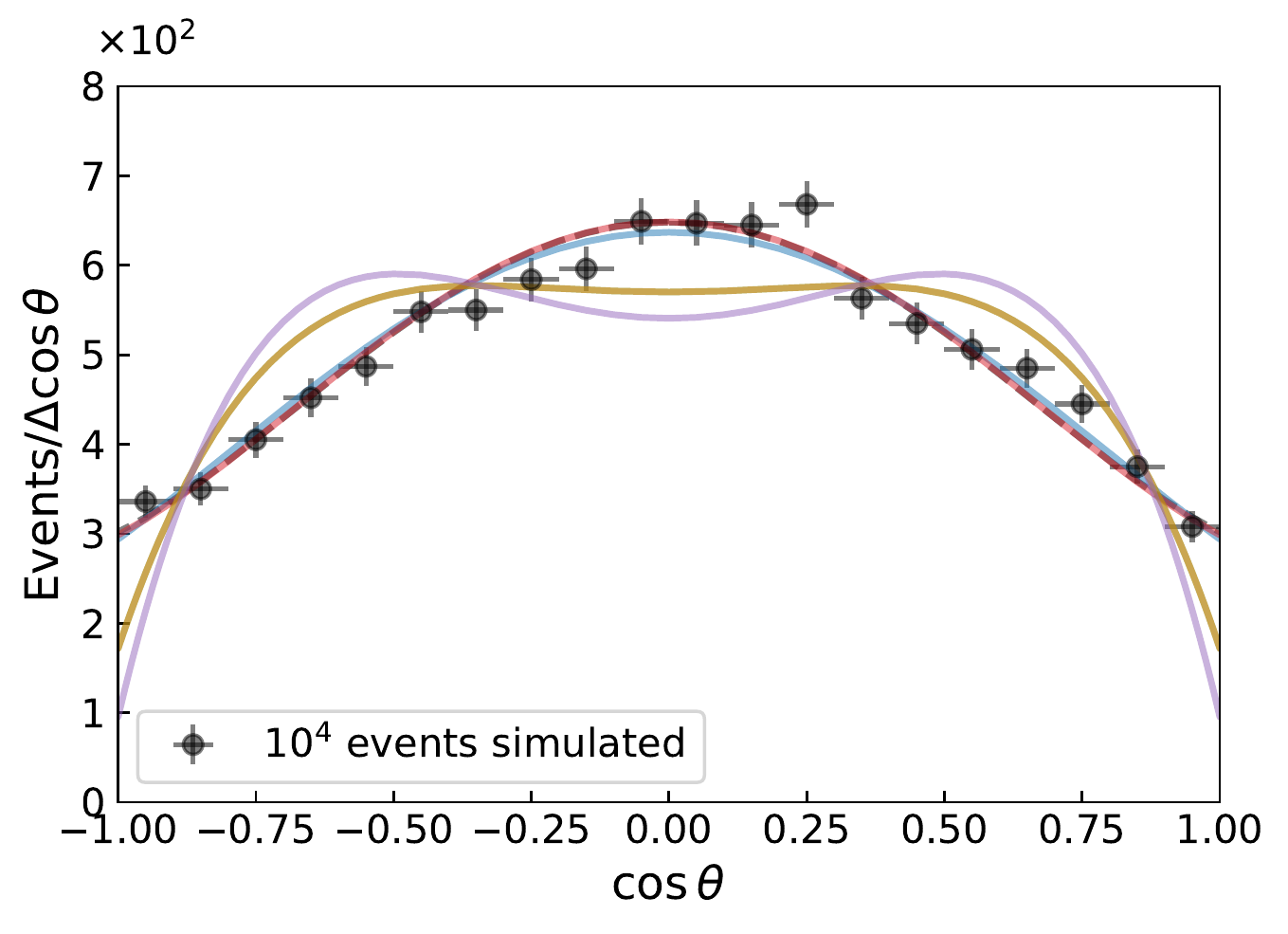} \\
    \includegraphics[width=0.9\columnwidth]{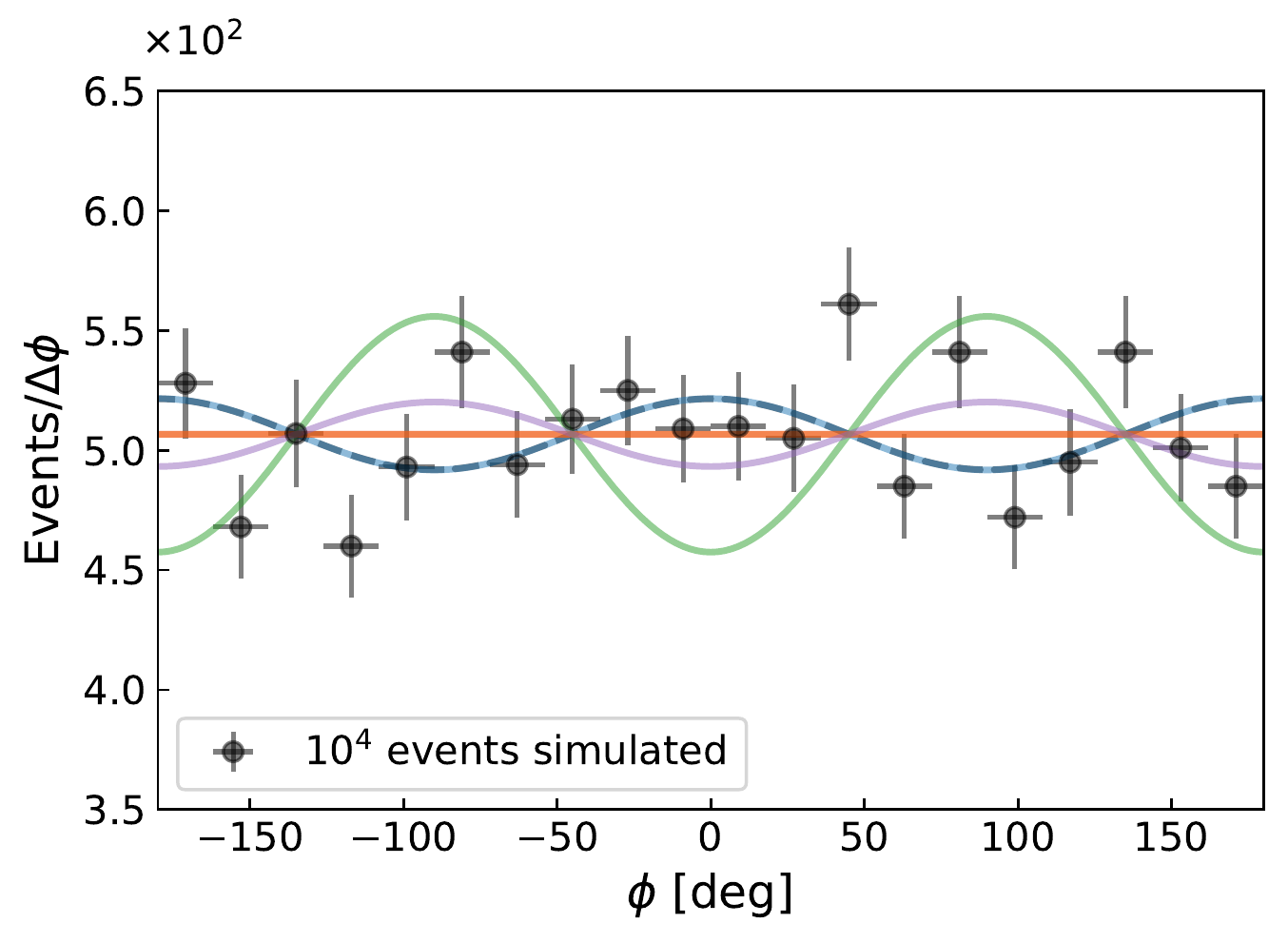} \\
    \includegraphics[width=0.9\columnwidth]{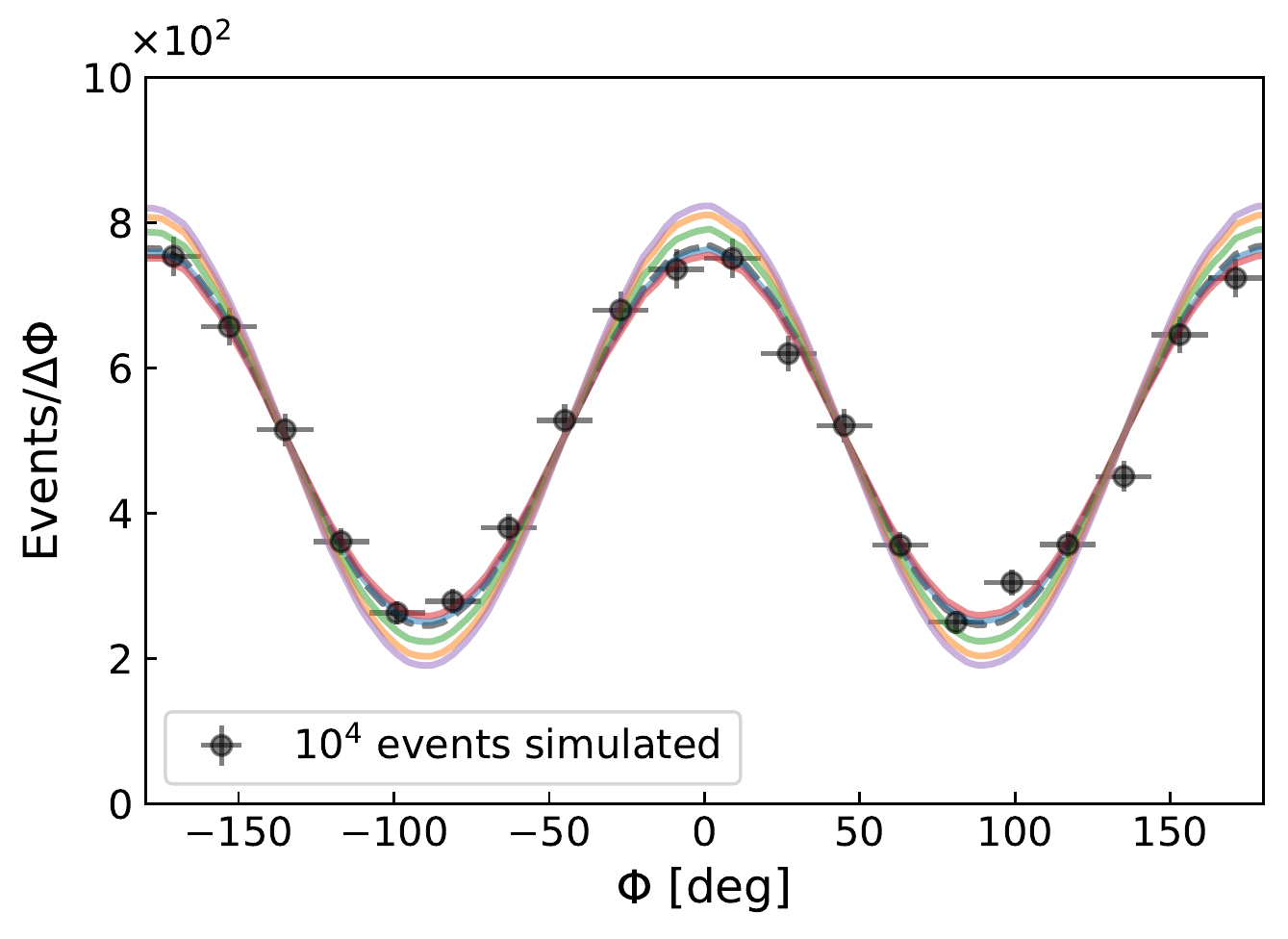}
    \end{tabular}
    \caption{\label{fig:sim_N10kplots} As \cref{fig:sim_N100plots} for fits to $10^4$ events with colors matching the plots in \cref{fig:sim_N10kwaves}.}
\end{figure}

\begin{figure}
    \centering
    \begin{tabular}{c}
    \includegraphics[width=0.9\columnwidth]{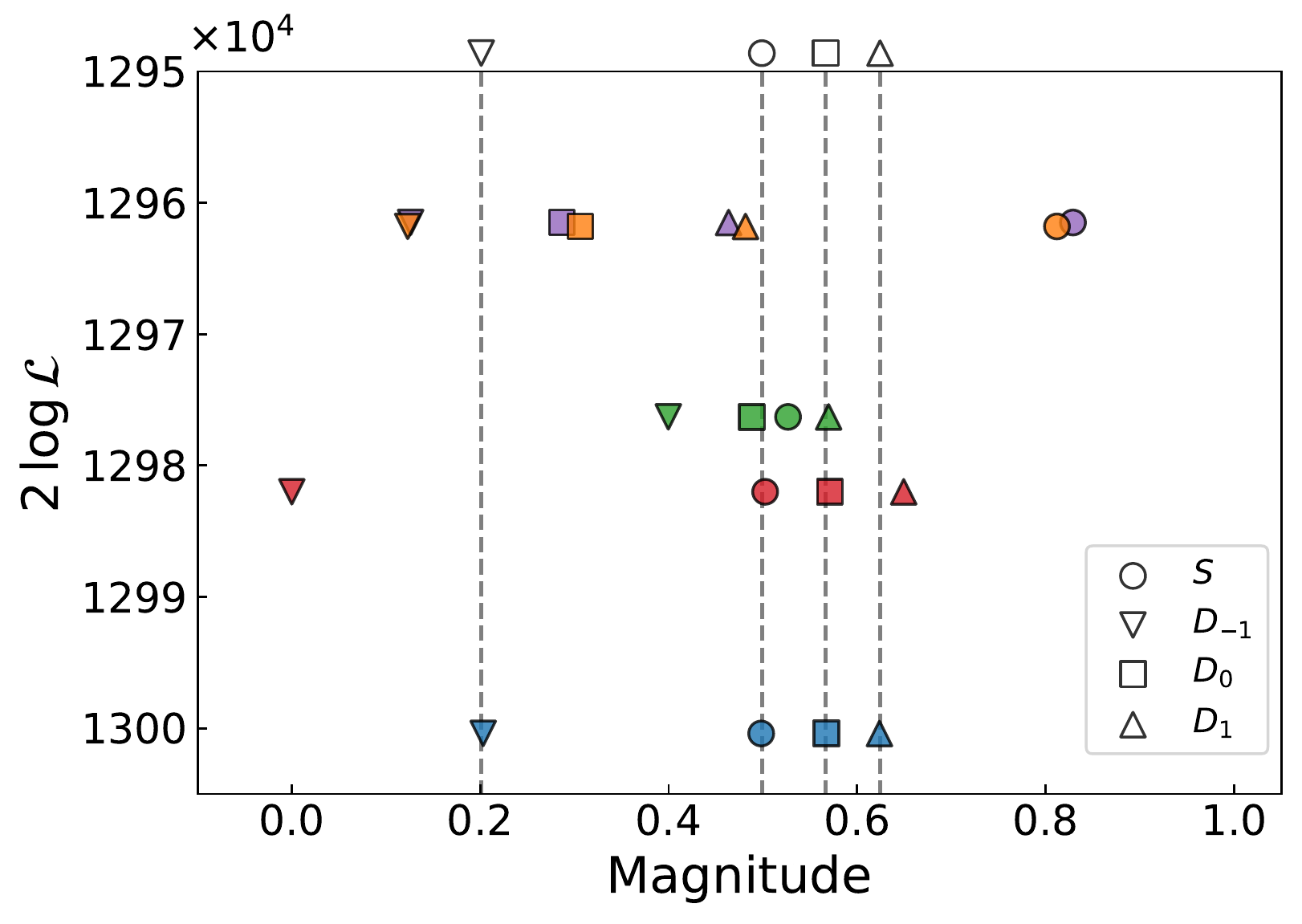} \\
    \includegraphics[width=0.9\columnwidth]{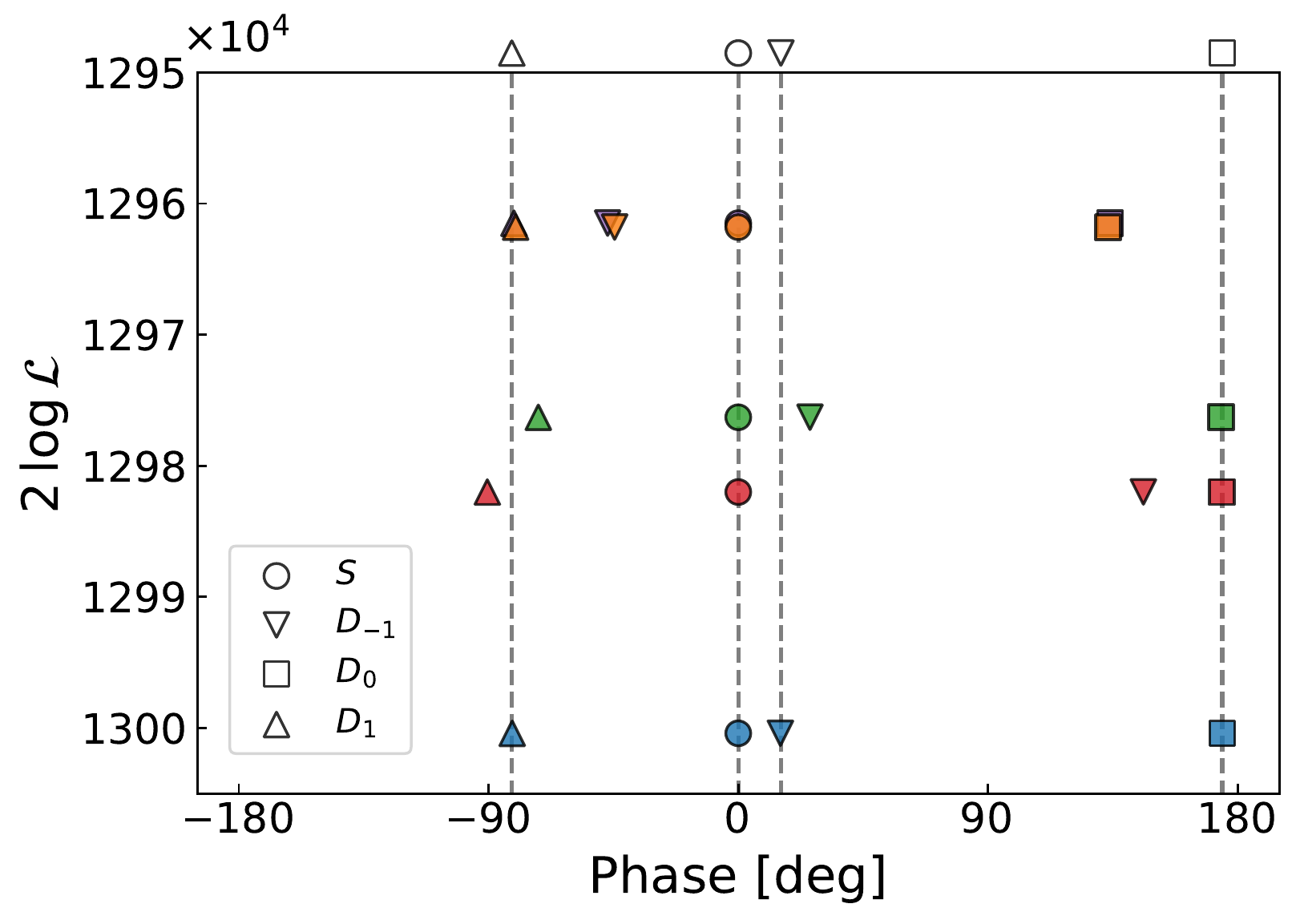} 
    \end{tabular}
    \caption{\label{fig:sim_N1Mwaves} As \cref{fig:sim_N100waves} for fits to $10^6$ events. Uncertainties are negligible and not shown.}
\end{figure}

\begin{figure}
    \centering
    \begin{tabular}{c}
    \includegraphics[width=0.9\columnwidth]{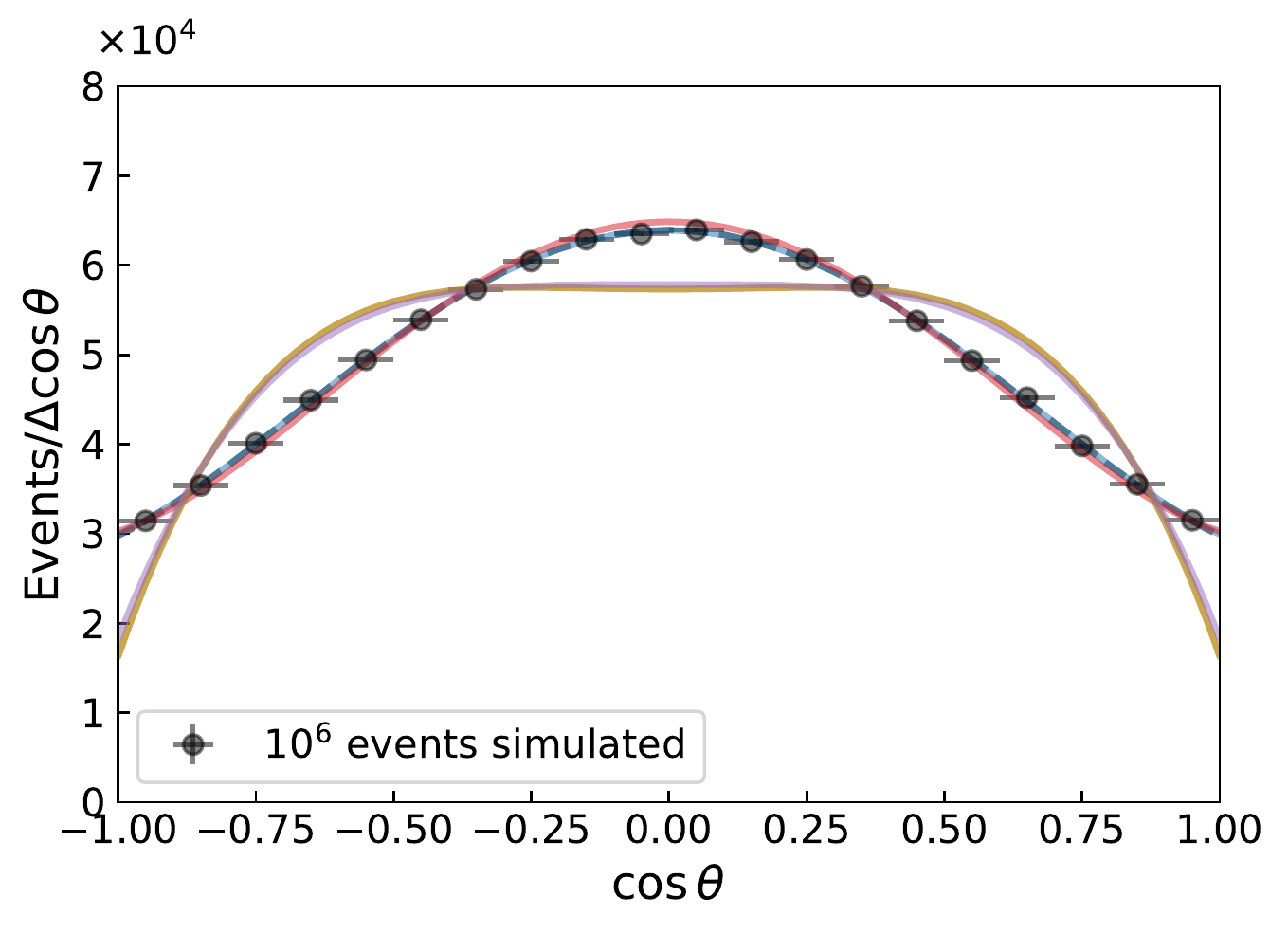} \\
    \includegraphics[width=0.9\columnwidth]{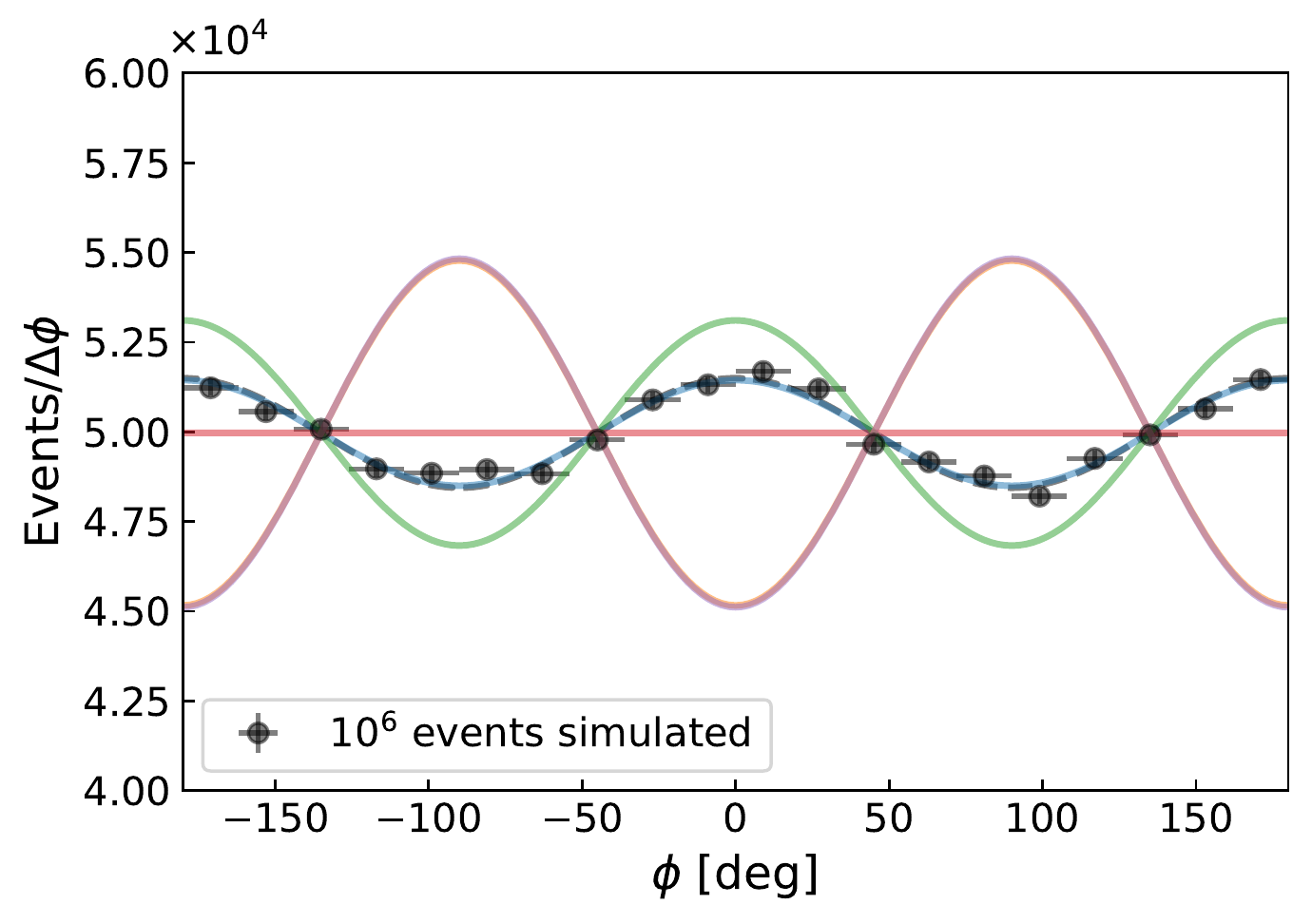} \\
    \includegraphics[width=0.9\columnwidth]{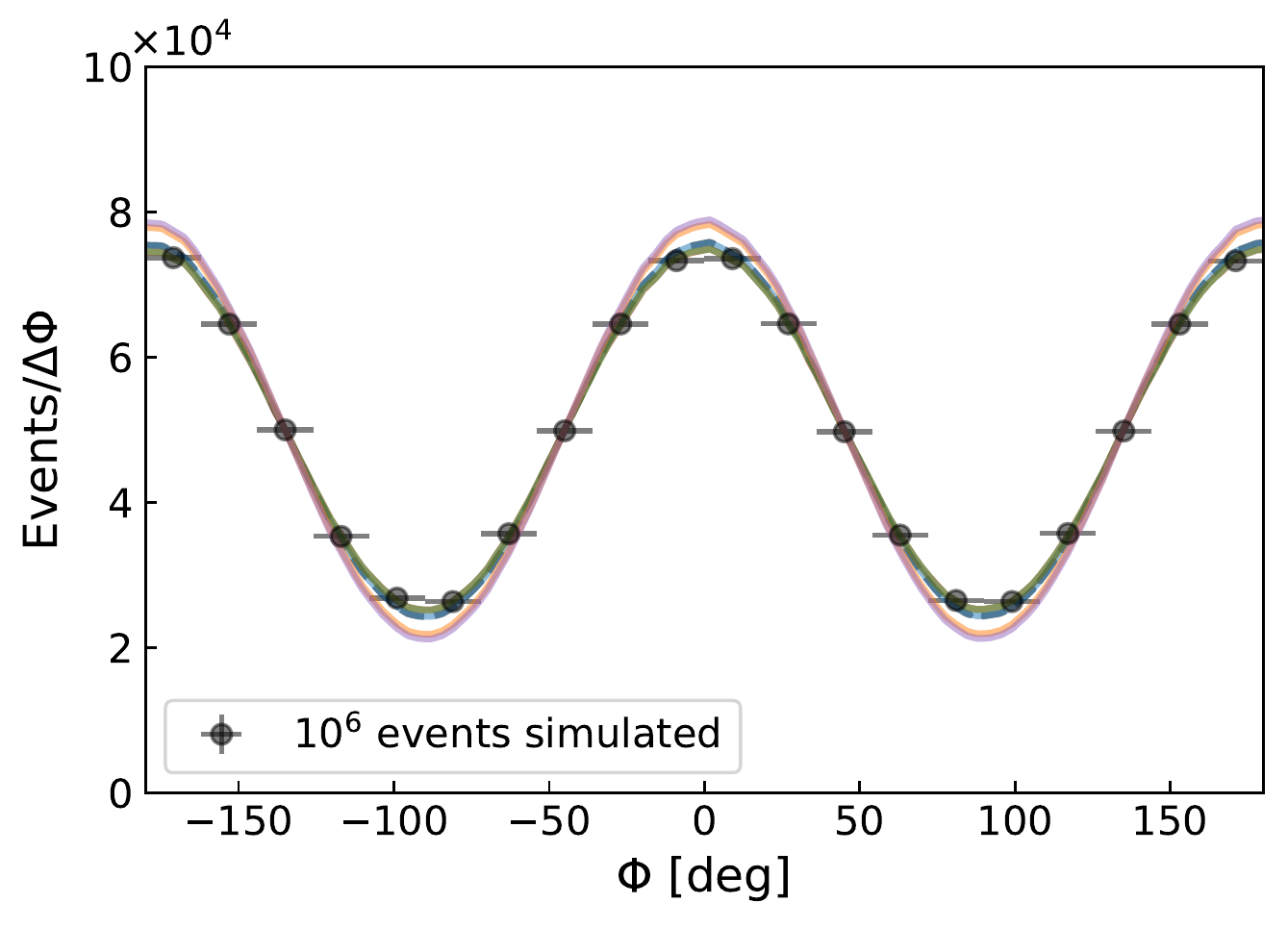}
    \end{tabular}
    \caption{\label{fig:sim_N1Mplots} As \cref{fig:sim_N100plots} for fits to $10^6$ events with colors matching the plots in \cref{fig:sim_N1Mwaves}.}
\end{figure}

We then performed event-by-event fits to extract
these four waves. We used \texttt{MINUIT}~\cite{James:1975dr} with random initial conditions to minimize the negative log likelihood ($-2 \log \cal{L}$). To explore the effect of differing statistical information on fit results, 
we examined three different cases with generated data sets of $10^2$, $10^4$, and $10^6$ events.

In \cref{fig:sim_N100waves} we show the resulting negative log likelihood and amplitude components from 50 fits to the pseudodata with 100 events. Similar results are shown for $10^4$ events in \cref{fig:sim_N10kwaves} and $10^6$ events in \cref{fig:sim_N1Mwaves}. For clarity, in these plots we show only a single complex conjugate solution set, though the fitting procedure did also identify trivial ambiguities, \ie the set with all phases simultaneously flipped in sign. 

In \cref{fig:sim_N100plots,fig:sim_N10kplots,fig:sim_N1Mplots} we show the projections of the intensity onto the polarization angle $\Phi$, and the decay angles $\phi$, $\theta$ for the best five solutions, compared to the distributions generated from the true amplitudes.

We observe in the case of the second best solution for each simulation (shown in red), the $\cos{\theta}$ distribution is almost identical to the best fit's distribution (blue).
However the second best $\phi$ distribution is flat (red), while the true distribution has a $\cos(2\phi)$ component. The reason for the flat distribution is that the magnitude of the $D_{-1}$ amplitude is zero for this solution (first red triangle in the plots of \cref{fig:sim_N100waves,fig:sim_N10kwaves,fig:sim_N1Mwaves}). 
The $h^0_2$ moment, which contributes to the $\cos(2\phi)$ amplitude, requires an interference between $D_{-1}$ and $D_{1}$, which is obviously zero when either of these waves has zero magnitude. Note that this solution is found despite \texttt{MINUIT} finishing with successful status.

The other solutions do not agree well with the data and therefore clearly do not represent real solutions, but rather represent artifacts of local maxima in the likelihood.
We also note that, for each level of statistics, we observe a similar behavior in the projection of the intensity onto $\Phi$ for all solutions. The best (blue) and second-best (red) solutions closely match the true solution (dashed black) for each case, while the less-favored solutions cannot be immediately discarded from this projection even at high statistics 

We should emphasize here that although we have shown explicitly that there are no mathematical ambiguities present, the false solutions found in fits to data or pseudodata must still be addressed. In fits to real data one may not always be able to extract the most favored solution from a fitting procedure and claim that it is the true, mathematically unique, solution due to detector effects and other systematics. In practice, each solution could be shifted up or down in likelihood, and the `true' solution could correspond to a local minimum rather than the global one. We do note that, in an environment with no systematics or detector effects, higher statistics allows one to make qualitative judgments about which solution best fits the data by considering projections of the intensity onto the scattering angles. We also note that in these simulations we have relatively few waves. In larger wave sets, the probability of finding the global minima from fifty random starting points reduces drastically. These issues are outside the scope of this paper, and we leave methods to address them for future work.

\section{Summary and Conclusions}
\label{sec:conclusions}
In this work, we have presented our formalism for the analysis of mathematical ambiguities for linearly polarized photoproduction of two spinless particles. We demonstrated for two wave sets that, even with a small number of constraints on the partial waves, the partial waves are over-specified by experimental data. We illustrated our results by generating pseudodata and extracting back the partial waves. We found that the best solution matches the input waves. 
We do not expect larger wave sets to exhibit root-conjugation ambiguities, as the number of constraints increases rapidly with the size of the fitted wave set. Rather, we expect that false solutions which appear in fits to real data come about as artifacts of complicated multidimensional properties of log-likelihood functions. These may be identified through examination of the angular dependence of the polarized observables. 

\acknowledgments
This work was supported by the U.S. Department of Energy contract DE-AC05-06OR23177, under which Jefferson Science Associates, LLC operates Jefferson Lab
and also by the U.S. Department of Energy Grant 
Nos.~DE-FG02-87ER40365, DE-FG02-92ER40735, and DE-FG02-87ER40315,
by the Spanish Ministerio de Ciencia e Innovaci\'on (MICINN) 
Grant Nos.~PID2019–106080GB-C21, PID2020-118758GB-I00, and PID2020-112777GB-I00,
and by the U.K.~Science and Technology Facilities Council under grants ST/P004458/1 and ST/V00106X/1.
VM is a Serra H\'unter fellow.
MA is supported by Generalitat Valenciana under Grant No.~CIDEGENT/2020/002.
CFR is supported by Spanish Ministerio de Educaci\'on y Formaci\'on Profesional (MEUFP) under Grant No.~BG20/00133.
The work of MM is funded by the Deutsche Forschungsgemeinschaft under Germany's Excellence Strategy-EXC-2094-390783311.
NH is supported by Polish research project Grant No.~2018/29/B/ST2/02576 (National Science Center). 
DW is supported by National Natural Science Foundation of China Grant No.~12035007 and the NSFC and the Deutsche Forschungsgemeinschaft (DFG, German Research Foundation) through the funds provided to the Sino-German Collaborative Research Center TRR110 ``Symmetries and the Emergence of Structure in QCD'' (NSFC Grant No.~12070131001, DFG ProjectID 196253076-TRR 110).
This research was supported by the Munich Institute for Astro-, Particle and BioPhysics (MIAPbP) which is funded by the Deutsche Forschungsgemeinschaft (DFG, German Research Foundation) under Germany's Excellence Strategy – EXC-2094 – 390783311. 
This work contributes to the aims of the U.S. Department of Energy ExoHad Topical Collaboration, contract DE-SC0023598. 

\appendix
\section{Barrelet zeros for spinless meson scattering} \label{sec:Barrelet}
We consider the elastic scattering of two spinless mesons~\cite{Barrelet:1971pw, Baker:1976xp}. Lorentz invariance allows us to choose the scattering plane as the $xz$ plane, and to write the intensity as a real positive function of the scattering angle $z=\cos\theta$. The differential cross section,
\begin{align}
    \frac{\diff \sigma}{\diff \Omega} & = |f(s, z)|^2,
\end{align}
is decomposed into partial waves of the decaying resonance, with angular momentum $\ell$ as 
\begin{align} \label{eq:pw2body}
    f(s,z) & = \sum_{\ell} (2\ell+1) a_\ell(s) P_\ell(z).
\end{align}
The center-of-mass energy $s$ is a fixed variable in our treatment. In practice, for each bin in $s$, the sum in \cref{eq:pw2body} is truncated to $\ell_\text{M}$ and the differential cross section is thus a polynomial of order $2\ell_M$ in the cosine of the scattering angle, $z$. The $\ell_\text{M}+1$ partial waves are in general complex numbers, but since the intensity is positive, the cross section can be factorized into its roots, also denoted Barrelet zeros~\cite{Barrelet:1971pw,Baker:1976xp}, in the following way
\begin{align}
    \frac{\diff \sigma}{\diff \Omega} & = C \prod_{i = 0}^{\ell_M} (z-z_i)(z-z_i^*),
\end{align}
where the $s$ dependence of the normalization factor $C$ and the Barrelet zeros $z_i$ have been omitted. 

Clearly, the knowledge of a set of partial waves $\{a_{\ell}\}$ determines the Barrelet zeros $\{z_i\}$, and {\it vice versa}. However, the differential cross section
includes both the roots $z_i$ and their conjugates $z_i^*$ while only one of $\{z_i,z_i^*\}$ is used to generate the partial waves; there is no physical distinction between a zero and its complex conjugate, which can lead to ambiguities in the values of the partial waves in \cref{eq:pw2body}. To see this, suppose
we know $\ell_M+1$ Barrelet zeros $\{z_i\}$ from which we reconstruct the partial waves:
\begin{align}
    a_\ell & = F_\ell(z_0,z_1,\ldots,z_{\ell_M-1} ,z_{\ell_M}),
\end{align}
where the functions $F_\ell$ are known for a given $\ell_M$. 
Alternatively one could choose to use the complex conjugate of any of the $\ell_M+1$ Barrelet zeros. For instance by choosing
\begin{align}
    a_\ell' & = F_\ell(z^*_0,z_1,\ldots,z^*_{\ell_M-1} ,z_{\ell_M}).
\end{align}
There are $2^{\ell_M+1}$ sets of potentially ambiguous partial waves $\{a_\ell'\}$ which lead to the same differential cross section. 
One can always rotate all the waves with a constant phase (in each bin of energy) such that the $S$-wave is real and positive. We are nevertheless left with $2^{\ell_M}$ possibilities for the partial waves in the case of spinless meson scattering. 

\bibliographystyle{apsrev4-1}
\bibliography{refs}

\begin{thebibliography}{20}%
\makeatletter
\providecommand \@ifxundefined [1]{%
 \@ifx{#1\undefined}
}%
\providecommand \@ifnum [1]{%
 \ifnum #1\expandafter \@firstoftwo
 \else \expandafter \@secondoftwo
 \fi
}%
\providecommand \@ifx [1]{%
 \ifx #1\expandafter \@firstoftwo
 \else \expandafter \@secondoftwo
 \fi
}%
\providecommand \natexlab [1]{#1}%
\providecommand \enquote  [1]{``#1''}%
\providecommand \bibnamefont  [1]{#1}%
\providecommand \bibfnamefont [1]{#1}%
\providecommand \citenamefont [1]{#1}%
\providecommand \href@noop [0]{\@secondoftwo}%
\providecommand \href [0]{\begingroup \@sanitize@url \@href}%
\providecommand \@href[1]{\@@startlink{#1}\@@href}%
\providecommand \@@href[1]{\endgroup#1\@@endlink}%
\providecommand \@sanitize@url [0]{\catcode `\\12\catcode `\$12\catcode
  `\&12\catcode `\#12\catcode `\^12\catcode `\_12\catcode `\%12\relax}%
\providecommand \@@startlink[1]{}%
\providecommand \@@endlink[0]{}%
\providecommand \url  [0]{\begingroup\@sanitize@url \@url }%
\providecommand \@url [1]{\endgroup\@href {#1}{\urlprefix }}%
\providecommand \urlprefix  [0]{URL }%
\providecommand \Eprint [0]{\href }%
\providecommand \doibase [0]{http://dx.doi.org/}%
\providecommand \selectlanguage [0]{\@gobble}%
\providecommand \bibinfo  [0]{\@secondoftwo}%
\providecommand \bibfield  [0]{\@secondoftwo}%
\providecommand \translation [1]{[#1]}%
\providecommand \BibitemOpen [0]{}%
\providecommand \bibitemStop [0]{}%
\providecommand \bibitemNoStop [0]{.\EOS\space}%
\providecommand \EOS [0]{\spacefactor3000\relax}%
\providecommand \BibitemShut  [1]{\csname bibitem#1\endcsname}%
\let\auto@bib@innerbib\@empty
\bibitem [{\citenamefont {Barrelet}(1972)}]{Barrelet:1971pw}%
  \BibitemOpen
  \bibfield  {author} {\bibinfo {author} {\bibfnamefont {E.}~\bibnamefont
  {Barrelet}},\ }\href {\doibase 10.1007/BF02732655} {\bibfield  {journal}
  {\bibinfo  {journal} {Nuovo Cim. A}\ }\textbf {\bibinfo {volume} {8}},\
  \bibinfo {pages} {331} (\bibinfo {year} {1972})}\BibitemShut {NoStop}%
\bibitem [{\citenamefont {Chung}(1997)}]{Chung:1997qd}%
  \BibitemOpen
  \bibfield  {author} {\bibinfo {author} {\bibfnamefont {S.~U.}\ \bibnamefont
  {Chung}},\ }\href {\doibase 10.1103/PhysRevD.56.7299} {\bibfield  {journal}
  {\bibinfo  {journal} {Phys. Rev. D}\ }\textbf {\bibinfo {volume} {56}},\
  \bibinfo {pages} {7299} (\bibinfo {year} {1997})}\BibitemShut {NoStop}%
\bibitem [{\citenamefont {Sadovsky}(1999)}]{Sadovsky:1999gt}%
  \BibitemOpen
  \bibfield  {author} {\bibinfo {author} {\bibfnamefont {S.~A.}\ \bibnamefont
  {Sadovsky}},\ }\href@noop {} {\bibfield  {journal} {\bibinfo  {journal}
  {Phys. Atom. Nucl.}\ }\textbf {\bibinfo {volume} {62}},\ \bibinfo {pages}
  {519} (\bibinfo {year} {1999})}\BibitemShut {NoStop}%
\bibitem [{\citenamefont {Austregesilo}(2014)}]{Austregesilo:2014oxa}%
  \BibitemOpen
  \bibfield  {author} {\bibinfo {author} {\bibfnamefont {A.}~\bibnamefont
  {Austregesilo}},\ }\emph {\bibinfo {title} {{Central Production of
  Two-Pseudoscalar Meson Systems at the COMPASS Experiment at CERN}}},\ \href
  {https://cds.cern.ch/record/1974100?ln=en} {Ph.D. thesis},\ \bibinfo
  {school} {Munich, Tech. U.} (\bibinfo {year} {2014})\BibitemShut {NoStop}%
\bibitem [{\citenamefont {Ketzer}\ \emph {et~al.}(2020)\citenamefont {Ketzer},
  \citenamefont {Grube},\ and\ \citenamefont {Ryabchikov}}]{Ketzer:2019wmd}%
  \BibitemOpen
  \bibfield  {author} {\bibinfo {author} {\bibfnamefont {B.}~\bibnamefont
  {Ketzer}}, \bibinfo {author} {\bibfnamefont {B.}~\bibnamefont {Grube}}, \
  and\ \bibinfo {author} {\bibfnamefont {D.}~\bibnamefont {Ryabchikov}},\
  }\href {\doibase 10.1016/j.ppnp.2020.103755} {\bibfield  {journal} {\bibinfo
  {journal} {Prog. Part. Nucl. Phys.}\ }\textbf {\bibinfo {volume} {113}},\
  \bibinfo {pages} {103755} (\bibinfo {year} {2020})},\ \Eprint
  {http://arxiv.org/abs/1909.06366} {arXiv:1909.06366 [hep-ex]} \BibitemShut
  {NoStop}%
\bibitem [{\citenamefont {Rodas}\ \emph {et~al.}(2022)\citenamefont {Rodas},
  \citenamefont {Pilloni}, \citenamefont {Albaladejo}, \citenamefont
  {Fern\'andez-Ram\'irez}, \citenamefont {Mathieu},\ and\ \citenamefont
  {Szczepaniak}}]{Rodas:2021tyb}%
  \BibitemOpen
  \bibfield  {author} {\bibinfo {author} {\bibfnamefont {A.}~\bibnamefont
  {Rodas}}, \bibinfo {author} {\bibfnamefont {A.}~\bibnamefont {Pilloni}},
  \bibinfo {author} {\bibfnamefont {M.}~\bibnamefont {Albaladejo}}, \bibinfo
  {author} {\bibfnamefont {C.}~\bibnamefont {Fern\'andez-Ram\'irez}}, \bibinfo
  {author} {\bibfnamefont {V.}~\bibnamefont {Mathieu}}, \ and\ \bibinfo
  {author} {\bibfnamefont {A.~P.}\ \bibnamefont {Szczepaniak}} (\bibinfo
  {collaboration} {JPAC}),\ }\href {\doibase 10.1140/epjc/s10052-022-10014-8}
  {\bibfield  {journal} {\bibinfo  {journal} {Eur. Phys. J. C}\ }\textbf
  {\bibinfo {volume} {82}},\ \bibinfo {pages} {80} (\bibinfo {year} {2022})},\
  \Eprint {http://arxiv.org/abs/2110.00027} {arXiv:2110.00027 [hep-ph]}
  \BibitemShut {NoStop}%
\bibitem [{\citenamefont {Gao}\ \emph {et~al.}(2023)\citenamefont {Gao},
  \citenamefont {Rong}, \citenamefont {Yang}, \citenamefont {Zhang},\ and\
  \citenamefont {Zhang}}]{Gao:2023jtq}%
  \BibitemOpen
  \bibfield  {author} {\bibinfo {author} {\bibfnamefont {Y.}~\bibnamefont
  {Gao}}, \bibinfo {author} {\bibfnamefont {T.}~\bibnamefont {Rong}}, \bibinfo
  {author} {\bibfnamefont {Z.}~\bibnamefont {Yang}}, \bibinfo {author}
  {\bibfnamefont {C.}~\bibnamefont {Zhang}}, \ and\ \bibinfo {author}
  {\bibfnamefont {Y.}~\bibnamefont {Zhang}},\ }\href@noop {} {\  (\bibinfo
  {year} {2023})},\ \Eprint {http://arxiv.org/abs/2302.13862} {arXiv:2302.13862
  [hep-ph]} \BibitemShut {NoStop}%
\bibitem [{\citenamefont {Kroenert}\ \emph {et~al.}(2023)\citenamefont
  {Kroenert}, \citenamefont {Wunderlich}, \citenamefont {Afzal},\ and\
  \citenamefont {Thiel}}]{Kroenert:2023ovd}%
  \BibitemOpen
  \bibfield  {author} {\bibinfo {author} {\bibfnamefont {P.}~\bibnamefont
  {Kroenert}}, \bibinfo {author} {\bibfnamefont {Y.}~\bibnamefont
  {Wunderlich}}, \bibinfo {author} {\bibfnamefont {F.}~\bibnamefont {Afzal}}, \
  and\ \bibinfo {author} {\bibfnamefont {A.}~\bibnamefont {Thiel}},\
  }\href@noop {} {\  (\bibinfo {year} {2023})},\ \Eprint
  {http://arxiv.org/abs/2305.10367} {arXiv:2305.10367 [nucl-th]} \BibitemShut
  {NoStop}%
\bibitem [{\citenamefont {Kok}\ and\ \citenamefont {{de
  Roo}}(1976)}]{Kok:1976}%
  \BibitemOpen
  \bibfield  {author} {\bibinfo {author} {\bibfnamefont {L.}~\bibnamefont
  {Kok}}\ and\ \bibinfo {author} {\bibfnamefont {M.}~\bibnamefont {{de Roo}}},\
  }\href {\doibase https://doi.org/10.1016/0550-3213(76)90479-X} {\bibfield
  {journal} {\bibinfo  {journal} {Nucl. Phys. B}\ }\textbf {\bibinfo {volume}
  {111}},\ \bibinfo {pages} {39} (\bibinfo {year} {1976})}\BibitemShut
  {NoStop}%
\bibitem [{\citenamefont {Ablikim}\ \emph {et~al.}(2015)\citenamefont {Ablikim}
  \emph {et~al.}}]{BESIII:2015rug}%
  \BibitemOpen
  \bibfield  {author} {\bibinfo {author} {\bibfnamefont {M.}~\bibnamefont
  {Ablikim}} \emph {et~al.} (\bibinfo {collaboration} {BESIII}),\ }\href
  {\doibase 10.1103/PhysRevD.92.052003} {\bibfield  {journal} {\bibinfo
  {journal} {Phys. Rev. D}\ }\textbf {\bibinfo {volume} {92}},\ \bibinfo
  {pages} {052003} (\bibinfo {year} {2015})},\ \bibinfo {note} {[Erratum:
  Phys.Rev.D 93, 039906 (2016)]},\ \Eprint {http://arxiv.org/abs/1506.00546}
  {arXiv:1506.00546 [hep-ex]} \BibitemShut {NoStop}%
\bibitem [{\citenamefont {Mathieu}\ \emph {et~al.}(2019)\citenamefont
  {Mathieu}, \citenamefont {Albaladejo}, \citenamefont
  {Fern\'andez-Ram\'\i{}rez}, \citenamefont {Jackura}, \citenamefont
  {Mikhasenko}, \citenamefont {Pilloni},\ and\ \citenamefont
  {Szczepaniak}}]{Mathieu:2019fts}%
  \BibitemOpen
  \bibfield  {author} {\bibinfo {author} {\bibfnamefont {V.}~\bibnamefont
  {Mathieu}}, \bibinfo {author} {\bibfnamefont {M.}~\bibnamefont {Albaladejo}},
  \bibinfo {author} {\bibfnamefont {C.}~\bibnamefont
  {Fern\'andez-Ram\'\i{}rez}}, \bibinfo {author} {\bibfnamefont {A.~W.}\
  \bibnamefont {Jackura}}, \bibinfo {author} {\bibfnamefont {M.}~\bibnamefont
  {Mikhasenko}}, \bibinfo {author} {\bibfnamefont {A.}~\bibnamefont {Pilloni}},
  \ and\ \bibinfo {author} {\bibfnamefont {A.~P.}\ \bibnamefont {Szczepaniak}}
  (\bibinfo {collaboration} {JPAC}),\ }\href {\doibase
  10.1103/PhysRevD.100.054017} {\bibfield  {journal} {\bibinfo  {journal}
  {Phys. Rev. D}\ }\textbf {\bibinfo {volume} {100}},\ \bibinfo {pages}
  {054017} (\bibinfo {year} {2019})},\ \Eprint
  {http://arxiv.org/abs/1906.04841} {arXiv:1906.04841 [hep-ph]} \BibitemShut
  {NoStop}%
\bibitem [{\citenamefont {Meyer}\ and\ \citenamefont
  {Van~Haarlem}(2010)}]{Meyer:2010ku}%
  \BibitemOpen
  \bibfield  {author} {\bibinfo {author} {\bibfnamefont {C.~A.}\ \bibnamefont
  {Meyer}}\ and\ \bibinfo {author} {\bibfnamefont {Y.}~\bibnamefont
  {Van~Haarlem}},\ }\href {\doibase 10.1103/PhysRevC.82.025208} {\bibfield
  {journal} {\bibinfo  {journal} {Phys. Rev. C}\ }\textbf {\bibinfo {volume}
  {82}},\ \bibinfo {pages} {025208} (\bibinfo {year} {2010})},\ \Eprint
  {http://arxiv.org/abs/1004.5516} {arXiv:1004.5516 [nucl-ex]} \BibitemShut
  {NoStop}%
\bibitem [{\citenamefont {Celentano}\ \emph {et~al.}(2020)\citenamefont
  {Celentano} \emph {et~al.}}]{CLAS:2020rdz}%
  \BibitemOpen
  \bibfield  {author} {\bibinfo {author} {\bibfnamefont {A.}~\bibnamefont
  {Celentano}} \emph {et~al.} (\bibinfo {collaboration} {CLAS}),\ }\href
  {\doibase 10.1103/PhysRevC.102.032201} {\bibfield  {journal} {\bibinfo
  {journal} {Phys. Rev. C}\ }\textbf {\bibinfo {volume} {102}},\ \bibinfo
  {pages} {032201} (\bibinfo {year} {2020})},\ \Eprint
  {http://arxiv.org/abs/2004.05359} {arXiv:2004.05359 [nucl-ex]} \BibitemShut
  {NoStop}%
\bibitem [{\citenamefont {Rodas}\ \emph {et~al.}(2019)\citenamefont {Rodas}
  \emph {et~al.}}]{JPAC:2018zyd}%
  \BibitemOpen
  \bibfield  {author} {\bibinfo {author} {\bibfnamefont {A.}~\bibnamefont
  {Rodas}} \emph {et~al.} (\bibinfo {collaboration} {JPAC}),\ }\href {\doibase
  10.1103/PhysRevLett.122.042002} {\bibfield  {journal} {\bibinfo  {journal}
  {Phys. Rev. Lett.}\ }\textbf {\bibinfo {volume} {122}},\ \bibinfo {pages}
  {042002} (\bibinfo {year} {2019})},\ \Eprint
  {http://arxiv.org/abs/1810.04171} {arXiv:1810.04171 [hep-ph]} \BibitemShut
  {NoStop}%
\bibitem [{\citenamefont {Kopf}\ \emph {et~al.}(2021)\citenamefont {Kopf},
  \citenamefont {Albrecht}, \citenamefont {Koch}, \citenamefont {K\"u\ss{}ner},
  \citenamefont {Pychy}, \citenamefont {Qin},\ and\ \citenamefont
  {Wiedner}}]{Kopf:2020yoa}%
  \BibitemOpen
  \bibfield  {author} {\bibinfo {author} {\bibfnamefont {B.}~\bibnamefont
  {Kopf}}, \bibinfo {author} {\bibfnamefont {M.}~\bibnamefont {Albrecht}},
  \bibinfo {author} {\bibfnamefont {H.}~\bibnamefont {Koch}}, \bibinfo {author}
  {\bibfnamefont {M.}~\bibnamefont {K\"u\ss{}ner}}, \bibinfo {author}
  {\bibfnamefont {J.}~\bibnamefont {Pychy}}, \bibinfo {author} {\bibfnamefont
  {X.}~\bibnamefont {Qin}}, \ and\ \bibinfo {author} {\bibfnamefont
  {U.}~\bibnamefont {Wiedner}},\ }\href {\doibase
  10.1140/epjc/s10052-021-09821-2} {\bibfield  {journal} {\bibinfo  {journal}
  {Eur. Phys. J. C}\ }\textbf {\bibinfo {volume} {81}},\ \bibinfo {pages}
  {1056} (\bibinfo {year} {2021})},\ \Eprint {http://arxiv.org/abs/2008.11566}
  {arXiv:2008.11566 [hep-ph]} \BibitemShut {NoStop}%
\bibitem [{\citenamefont {Woss}\ \emph {et~al.}(2021)\citenamefont {Woss},
  \citenamefont {Dudek}, \citenamefont {Edwards}, \citenamefont {Thomas},\ and\
  \citenamefont {Wilson}}]{Woss:2020ayi}%
  \BibitemOpen
  \bibfield  {author} {\bibinfo {author} {\bibfnamefont {A.~J.}\ \bibnamefont
  {Woss}}, \bibinfo {author} {\bibfnamefont {J.~J.}\ \bibnamefont {Dudek}},
  \bibinfo {author} {\bibfnamefont {R.~G.}\ \bibnamefont {Edwards}}, \bibinfo
  {author} {\bibfnamefont {C.~E.}\ \bibnamefont {Thomas}}, \ and\ \bibinfo
  {author} {\bibfnamefont {D.~J.}\ \bibnamefont {Wilson}} (\bibinfo
  {collaboration} {Hadron Spectrum}),\ }\href {\doibase
  10.1103/PhysRevD.103.054502} {\bibfield  {journal} {\bibinfo  {journal}
  {Phys. Rev. D}\ }\textbf {\bibinfo {volume} {103}},\ \bibinfo {pages}
  {054502} (\bibinfo {year} {2021})},\ \Eprint
  {http://arxiv.org/abs/2009.10034} {arXiv:2009.10034 [hep-lat]} \BibitemShut
  {NoStop}%
\bibitem [{\citenamefont {Mathieu}\ \emph {et~al.}(2020)\citenamefont
  {Mathieu}, \citenamefont {Pilloni}, \citenamefont {Albaladejo}, \citenamefont
  {Bibrzycki}, \citenamefont {Celentano}, \citenamefont
  {Fern\'andez-Ram\'\i{}rez},\ and\ \citenamefont
  {Szczepaniak}}]{Mathieu:2020zpm}%
  \BibitemOpen
  \bibfield  {author} {\bibinfo {author} {\bibfnamefont {V.}~\bibnamefont
  {Mathieu}}, \bibinfo {author} {\bibfnamefont {A.}~\bibnamefont {Pilloni}},
  \bibinfo {author} {\bibfnamefont {M.}~\bibnamefont {Albaladejo}}, \bibinfo
  {author} {\bibfnamefont {L.}~\bibnamefont {Bibrzycki}}, \bibinfo {author}
  {\bibfnamefont {A.}~\bibnamefont {Celentano}}, \bibinfo {author}
  {\bibfnamefont {C.}~\bibnamefont {Fern\'andez-Ram\'\i{}rez}}, \ and\ \bibinfo
  {author} {\bibfnamefont {A.~P.}\ \bibnamefont {Szczepaniak}} (\bibinfo
  {collaboration} {JPAC}),\ }\href {\doibase 10.1103/PhysRevD.102.014003}
  {\bibfield  {journal} {\bibinfo  {journal} {Phys. Rev. D}\ }\textbf {\bibinfo
  {volume} {102}},\ \bibinfo {pages} {014003} (\bibinfo {year} {2020})},\
  \Eprint {http://arxiv.org/abs/2005.01617} {arXiv:2005.01617 [hep-ph]}
  \BibitemShut {NoStop}%
\bibitem [{\citenamefont {Adolph}\ \emph {et~al.}(2015)\citenamefont {Adolph}
  \emph {et~al.}}]{COMPASS:2014vkj}%
  \BibitemOpen
  \bibfield  {author} {\bibinfo {author} {\bibfnamefont {C.}~\bibnamefont
  {Adolph}} \emph {et~al.} (\bibinfo {collaboration} {COMPASS}),\ }\href
  {\doibase 10.1016/j.physletb.2014.11.058} {\bibfield  {journal} {\bibinfo
  {journal} {Phys. Lett. B}\ }\textbf {\bibinfo {volume} {740}},\ \bibinfo
  {pages} {303} (\bibinfo {year} {2015})},\ \bibinfo {note} {[Erratum:
  Phys.Lett.B 811, 135913 (2020)]},\ \Eprint {http://arxiv.org/abs/1408.4286}
  {arXiv:1408.4286 [hep-ex]} \BibitemShut {NoStop}%
\bibitem [{\citenamefont {James}\ and\ \citenamefont
  {Roos}(1975)}]{James:1975dr}%
  \BibitemOpen
  \bibfield  {author} {\bibinfo {author} {\bibfnamefont {F.}~\bibnamefont
  {James}}\ and\ \bibinfo {author} {\bibfnamefont {M.}~\bibnamefont {Roos}},\
  }\href {\doibase 10.1016/0010-4655(75)90039-9} {\bibfield  {journal}
  {\bibinfo  {journal} {Comput. Phys. Commun.}\ }\textbf {\bibinfo {volume}
  {10}},\ \bibinfo {pages} {343} (\bibinfo {year} {1975})}\BibitemShut
  {NoStop}%
\bibitem [{\citenamefont {Baker}(1976)}]{Baker:1976xp}%
  \BibitemOpen
  \bibfield  {author} {\bibinfo {author} {\bibfnamefont {R.~D.}\ \bibnamefont
  {Baker}},\ }\href
  {https://lib-extopc.kek.jp/preprints/PDF/1976/7606/7606017.pdf} {\enquote
  {\bibinfo {title} {{Barrelet Zeros in Partial Wave Analysis}},}\ } (\bibinfo
  {year} {1976})\BibitemShut {NoStop}%
\end{thebibliography}%
\end{document}